\documentclass{ws-ijmpd}
\usepackage{url}
\usepackage[super,compress]{cite}

\newcommand{\lsim}{\raisebox{-0.13cm}{~\shortstack{$<$ \\[-0.07cm] $\sim$}}~}
\newcommand{\gsim}{\raisebox{-0.13cm}{~\shortstack{$>$ \\[-0.07cm] $\sim$}}~}

\begin{document}

\markboth{Caputi}
{AGN and their Role in Galaxy Evolution: the Infrared Perspective}

%
\catchline{}{}{}{}{}
%

\title{ACTIVE GALACTIC NUCLEI AND THEIR ROLE IN GALAXY EVOLUTION: THE INFRARED PERSPECTIVE}

\author{K. I. CAPUTI}

\address{Kapteyn Astronomical Institute, University of Groningen, \\
P.O. Box 800, 9700 AV Groningen, The Netherlands\\
karina@astro.rug.nl}

\maketitle


\begin{abstract}
The remarkable progress made in infrared (IR) astronomical instruments over the last 10-15 years has radically changed our vision of the extragalactic IR sky, and overall understanding of galaxy evolution. In particular, this has been the case for the study of active galactic nuclei (AGN), for which IR observations provide a wealth of complementary information that cannot be derived from data in other wavelength regimes. In this review, I summarize the unique contribution that IR astronomy has recently made to our understanding of AGN and their role in galaxy evolution,  including both physical studies of AGN at IR wavelengths, and the search for AGN among IR galaxies in general. Finally, I identify and discuss key open issues that it should be possible to address with forthcoming IR telescopes.

\end{abstract}

\keywords{active galactic nuclei; infrared galaxies; galaxy evolution}

\ccode{PACS numbers: 98.54-h -- 98.58.Jg --  98.54.Ep -- 98.62.Js -- 98.62.Nx}

\section{Introduction}

The infrared (IR) spectral range (referring to wavelengths $\lambda \sim 1-1000 \, \rm \mu m$) contains a wealth of diagnostic features that are crucial to understand the physics of active galactic nuclei  (AGN) and their host galaxies. Over the last 10-15 years, enormous progress has been made in this field thanks to the advent of major IR space telescopes, as well as the increasing possibility of conducting high spatial resolution studies with IR telescopes from the ground. It is now commonly agreed that IR observations offer a unique view to the study of AGN,  which is complementary to studies conducted in other parts of the electromagnetic spectrum.

In a cosmological context, IR observations are essential to unveil the fundamental role of AGN in galaxy evolution. In the past, the most powerful nuclear and star formation activity of the Universe was obscured by dust, thus making the IR to be the primary regime to conduct a census of galaxy activity in the first half of cosmic time. Until only two decades ago, our vision of the IR Universe was basically limited to our local surroundings, so the open up of the IR Universe to very high redshifts, as it has been made possible over the last years, is a truly remarkable achievement.

The aim of this review is summarizing the most important IR studies of AGN of the last 10-15 years, including both physical studies of AGN at IR wavelengths, and the search for AGN among IR galaxies in general. The results discussed here correspond almost exclusively  to near-IR through millimetre studies, but a few references to results beyond these wavelength regimes  are included when appropriate. The breadth of the compiled literature should hopefully help the reader find recent, relevant references for the main topics related to AGN studies at IR wavelengths. The layout of this paper is as follows. In Section \S\ref{sec_agnprop}, I analyse the main AGN properties determined through IR observations, including evidence for the existence and characteristics of the dusty torus, the circumnuclear region, and the AGN host galaxies. Sections \S\ref{sec_agnsfr} and \S\ref{sec_feed} deal with the AGN/star formation (SF) connection, and evidence for  AGN feedback. In section \S\ref{sec_stat}, I review different methods to identify AGN in IR galaxy samples, and general statistical studies of AGN through cosmic time. Finally, in Section \S\ref{sec_fut}, I  briefly discuss some of the main open questions in the field,  and the possibility of addressing them with future IR telescopes.

\section{AGN Properties Determined through IR Observations}
\label{sec_agnprop}

\subsection{Studies of the Dusty Torus}
\label{sec_torus}

The presence of a dusty structure with a torus-like geometry surrounding AGN at the parsec (pc) scales
was first inferred indirectly by early optical spectropolarimetric observations\cite{ant85}.
The detection of broad permitted lines (in particular H$\beta$) in the spectrum
of type 2 AGN, strongly suggested that these nuclei are intrinsically
similar to type 1 AGN, but obscured along our line of sight by a dusty medium on small scales -- small
enough not to obscure the narrow-line region (NLR), but large enough to obscure the broad line region (BLR). At that
time, a torus-like geometry was suggested to break the axisymmetry along the
line of sight, and hence produce the observed polarisation of the scattered
emission lines. This was the initial result pointing towards a `Unified Model' for AGN.

The detection of deep near/mid-IR absorption features, and in particular
the 9.7$\mu$m silicate absorption, typically observed in the mid-IR spectra of type 2 AGN, has provided strong indication for the presence of obscuring dusty material along the line of sight\cite{sie05,shi06,hao07,net07}. The depth of the silicate
feature is observed to correlate (although with large scatter) with the gaseous
column density observed in the X-rays, indicating that generally the gaseous
and dusty absorbers are generally co-planar. One should recall here that the
studies on the variability of the absorbing column density in the X-ray has revealed
that the bulk of the absorbing gas (especially at column densities $\rm
N_H>10^{23}~cm^{-2}$) is actually within the BLR, i.e. within the
dust sublimation radius\cite{ris02,elv04}, and therefore it was not obvious
that dust absorption should correlate with gaseous absorption, because at least
part of the latter occurs on much smaller scales.
This large scatter, and the fact that the silicate feature is observed also in emission (typically in Seyferts 1, but also in a few Seyferts 2), has been interpreted in the context of a clumpy absorbing medium\cite{shi06}.

Direct detection of the putative dusty torus (in emission) has
been more difficult. The presence of strong near-IR and mid-IR emission (often
unresolved or barely resolved) in ground-based, seeing-limited observations have
clearly revealed the presence of dust close to the AGN\cite{kra01,alo01}.
This dust mainly emits at wavelengths around $\lambda\sim 2 \, \rm \mu m$, and the temperature inferred from such near/mid-IR component
is of the order of several hundred K.  From simple arguments of dust thermal equilibrium, it is inferred that this dust 
must be located at a distance of a few pc at most from the central engine. Indeed, to emit
at such short wavelengths, dust must be close to its sublimation temperature ($T\sim 1000-2000~K$). Dust sublimation is likely responsible for defining the inner edge of the putative absorbing torus, and also setting the outer edge of
the BLR. This has been tested through near-IR reverberation studies\cite{sug06,kis07}, which have shown that the location of the dust emitting at 2~$\mu$m is at a distance 0.01-0.3~pc, and scales with the AGN luminosity as $L^{1/2}$\cite{kis11}.

While it has been agreed that the inner edge of the torus is set by the
sublimation radius, the structure, geometry and extension of the dusty torus
has been much more difficult to characterise. Initially, the properties and geometry
of the torus were inferred by modelling its IR spectral energy distribution (SED), but the
obvious issue was the difficulty of disentangling the IR emission associated
with the AGN-heated dust, from the IR light emitted by the host galaxy. A major step forward in understanding the properties and geometry of the torus has been made thanks to the development of interferometric techniques at mid-IR wavelengths.
Real 2D maps have only been obtained for a handful of bright AGN, but at least partial information on the $uv$ plane has been obtained for many other cases\cite{jaf04,tri07,bur09,rab09,tri09,tri14}.

\begin{figure}[h]
\centerline{\psfig{file=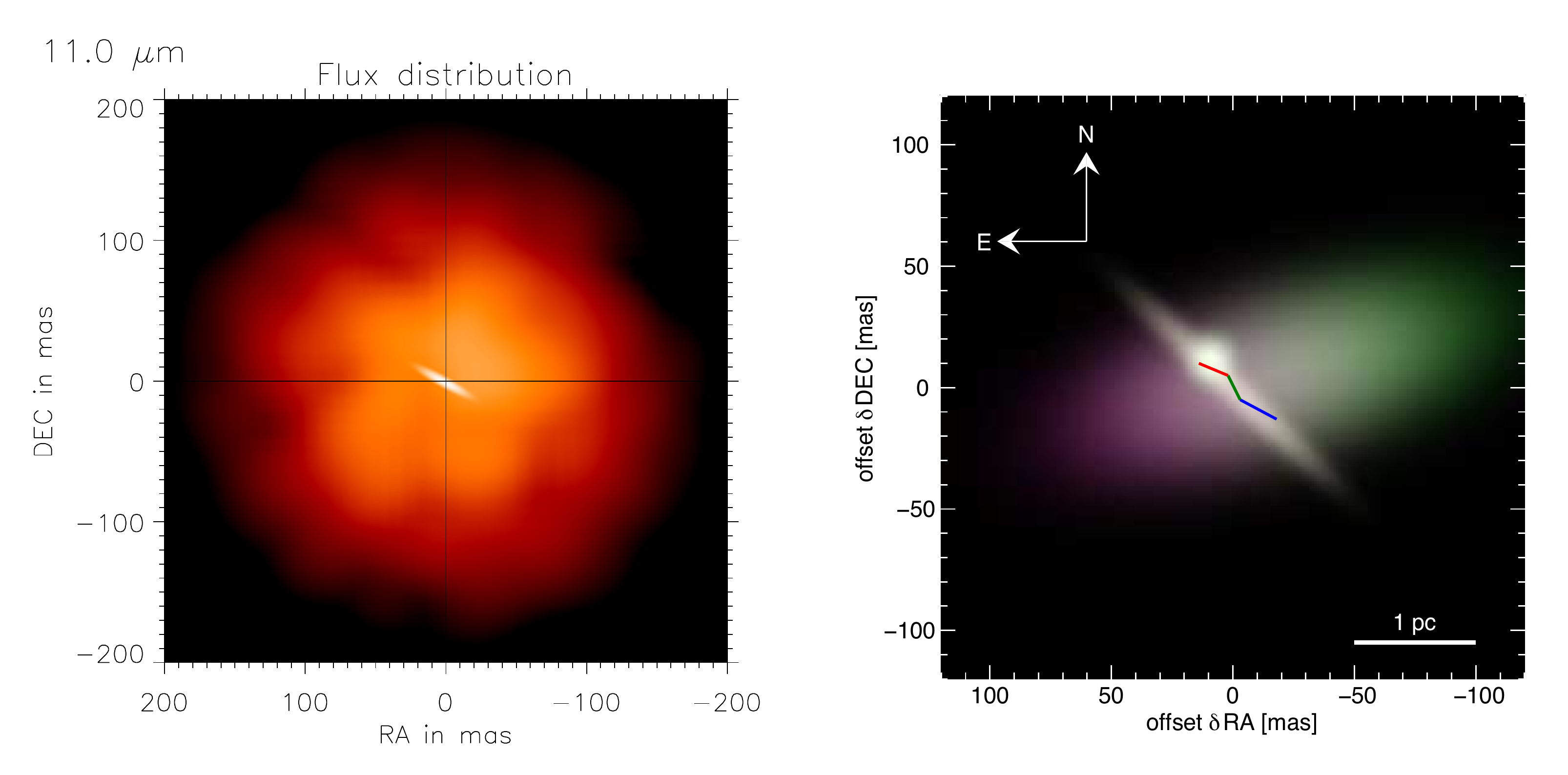,width=12cm}}
\caption{Distribution of the dusty medium in the nuclear region of the Circinus galaxy as inferred from mid-IR interferometric
observations. {\em Left:} flux distribution inferred from the mid-IR observations, with a very compact
sub-parsec scale elongated structure (the torus) and diffuse emission on the scale of 1-2 pc.  {\em Right:}  inferred
distribution of dust in the central region, in which the colours blue, green and red correspond to the model at 8.0, 10.5
and 13.0~$\rm \mu m$, respectively. Figures taken from Tristram et al. (2007, 2014)\cite{tri07,tri14}. Reproduced with permission from Astronomy \& Astrophysics, \copyright ESO.
\label{fig_circ}
}
\end{figure}

The prototypical Seyfert 2 galaxies NGC~1068 and Circinus are the two cases best studied with mid-IR interferometry.
The results for NGC1068 are consistent with a two-component dust distribution: an inner, 0.5~pc thick, and rather
elongated hot (T $>$ 800 K) component; and a more extended (3--4 pc), less elongated, colder (T $\sim$ 300 K) component. Most of the absorption appears to be located outside 1~pc. A similar result has been found for Circinus:  two components, one inner and more compact (0.4 pc), and another outer (2 pc) component (Fig.~\ref{fig_circ}). However, the temperature of the inner component in Circinus ($T = 330 \, \rm K$) is significantly lower than that in NGC 1068, and far from the sublimation temperature. Observations carried on the type 1 objects have provided results in agreement with those for Seyfert 2\cite{bur09,kis09}.

The main difficulty in reproducing the nuclear IR SED of AGN has been the broad distribution of the IR bump at $\lambda \sim 100-200 \, \rm \mu m$, which implies a wide range of temperatures. Initial models\cite{pie93} envisaging a compact pc-scale torus had serious problems in matching the observed IR SED, and also issues associated with hydrodynamical stability (although more recently Krolik (Ref.~\refcite{kro07}) pointed out that internal IR radiation can provide the required pressure for stability).  However, more recently it has been proposed that one of the main
problems of early studies has been the assumption that the torus must have a uniform density distribution (or at most
a smooth radial gradient). A clumpy torus, beside being physically more realistic, has the advantage that, if the dusty clouds are optically thick to the incoming radiation, then each individual small clump will have a range of temperatures,  and therefore can reproduce the observed broad SED\cite{nen02,eli06,sch05,hon06,mor09,ram09,alo11,ram11,sta12,lir13}. The overall distribution of clouds can be confined within a relatively small radius (equivalent to $\sim 30$ times the sublimation radius). The stability of the torus structure is in this case mostly given by the velocity dispersion of the clouds, through thermal and radiation pressure, as well as turbulence induced by supernovae and nuclear stellar winds which can help the torus  `puffing up'\cite{wad02,sch05}.

Interferometric studies on a sizeable sample of objects in the near- and mid-infrared have shown that no significant differences are found between type 1 and 2 sources. The torus sizes inferred from mid-IR interferometric observations tend to favor the clumpy, compact models. However, a definitive proof of these models can only come from sensitive and high angular resolution observations at sub-millimetre wavelengths, which will now become feasible with the Atacama Large Millimetre Array (ALMA) for the first time. Indeed, while the difference between clumpy and diffuse-extended models is minimal at mid-IR wavelength, the difference is much more dramatic at far-IR and sub-millimetre wavelengths, with the former predicting much smaller sizes than the latter.

\subsection{The Circumnuclear Region} 

Both the capability of piercing much deeper into the dust (relative to optical observations), and the access to transitions
associated with a broader range of ionisation and critical densities, has enabled IR spectroscopic studies
to characterise in detail the properties of the circumnuclear region of AGN. 
IR coronal lines are generally significantly broader than other lines observed in the narrow line region, although not as broad as the permitted lines coming from the broad line region, suggesting that they may be tracing an `intermediate line region', bridging the gap
between the BLR and the NLR, possibly including the inner surface of the obscuring torus.
This is supported by the imaging of the near-IR coronal lines, revealing that the emission is indeed much more compact
than the narrow lines traced in the optical and by lower ionisation species\cite{msa11}.

More generally, by combining the information from multiple IR emission lines, astronomers could model in detail
the physics of the circumnuclear region ionised by the AGN, constraining the ionisation parameter, gas density,
and also the shape of the ionising radiation spectrum\cite{spi05,dud07}. Near-IR integral-field spectroscopy of high ionisation lines have also enabled the detection of counter-ionisation cones, which where not seen at optical wavelengths, hence revealing that generally the one-sided ionisation cones at optical wavelengths are not due to asymmetric ionisation, but to dust obscuration in the galactic disk\cite{pri05,msa11}. 

Sensitive sub-/millimetre spectroscopy has recently provided new tools to investigate the circumnuclear region
of AGN. In particular, in the presence of an AGN, various molecular transitions are powerful tracers of the gas
excited by X-ray radiation, i.e. the so-called X-ray Dominated Regions (XDR).
Indeed, the hard X-rays typically produced by AGN can penetrate deep into the circumnuclear
medium, and they can heat it mainly through photoionisation of the dust grains. The X-ray enhanced chemistry favours the formation of specific molecular species and boosts the emission of specific molecular transitions, which can be used as a diagnostic of X-ray irradiated gas in the AGN circumnuclear region\cite{mal96,lep96,mei05,mei07}. For example, highly excited CO rotational lines can be a tracer of warm gas heated by X-rays\cite{spa08}. The detection of CO emission from very high $J$ transitions in nearby powerful AGN\cite{vdw10} has been regarded as evidence for extensive XDR. However, high-$J$ transitions are also excited by shocks\cite{pap10}. These are also expected in AGN as a consequence of the powerful winds that are generated in these objects, but also in powerful star-forming galaxies. Instead, the enhanced HCN(1-0) millimetre
emission in the central region of nearby AGN\cite{koh08,sani12}  has been regarded as more reliable evidence for X-ray irradiated gas. This is in principle a powerful tool to detect heavily obscured AGN, especially with ALMA.

\subsection{The AGN Host Galaxies}
\label{sec_hosts}

This section is mostly focused on the morphological and dynamical properties of AGN host galaxies,
while star formation in AGN hosts is discussed in section \S\ref{sec_agnsfr}.

Near-IR images are considered the optimal tool to investigate the morphological properties of
galaxies, and in particular AGN host galaxies, because the near-IR is a good tracer of a galaxy old stellar populations, and is less affected by recent star formation and dust extinction. Near-IR imaging surveys of local galaxies have not found any clear evidence that the morphology of local Seyfert galaxies is different, in terms of bar occurrence or morphological
distortions, with respect to galaxies not hosting AGN\cite{hun04}. This is not a surprising result, as the accretion
rate needed to fuel low luminosity AGN is so low ($\rm \dot{M}\sim 10^{-3}~M_{\odot}~yr^{-1}$) that
a single giant molecular cloud located at the center of the galaxy is enough to accrete the black hole over
a significant fraction of the Hubble time, without the need of supplying additional gas from the host
galaxy through non-axisymmetric potentials.

Studies of the gas dynamics through interferometric mapping of the molecular gas in galaxies hosting nearby
AGN have provided consistent results. The dynamics of the molecular gas does not show evidence for
an excess of streaming motions towards the nucleus in nearby
galaxies hosting low luminosity AGN\cite{gb07}. However, what can be more important for low luminosity AGNs is the circumnuclear dynamics, on small scales of $\sim$100~pc. Different IR/millimetre, as well as optical, observations have revealed that low-luminosity AGN tend to be associated with spiral structures at those small scales\cite{sto07,rif08}.

\begin{figure}[!h]
\includegraphics[scale=0.25]{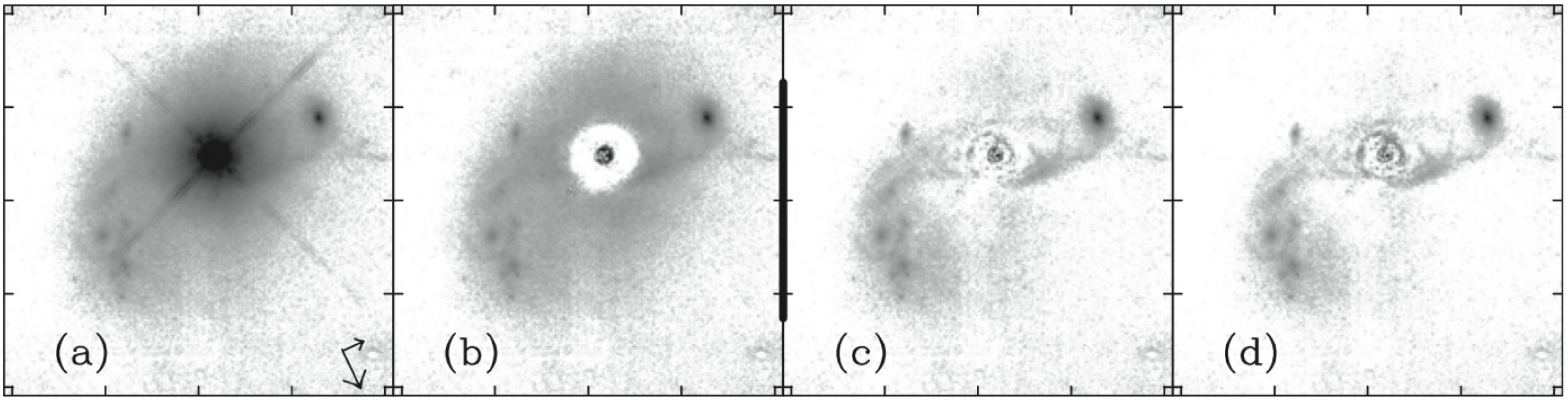}
\caption{a) Near-IR {\em HST} image of a nearby quasar. The subtraction of the
central point spread function (PSF) associated with the quasar point-like source (panels b to d), reveals a strongly interacting
system with prominent tidal features. Figure taken from Veilleux et al.~(2009)\cite{vei09}. \copyright AAS. Reproduced with permission. \label{fig_pgmorph}}
\end{figure}

The situation is quite different for the most luminous AGN, particularly those found in local ultra-luminous infrared galaxies (ULIRGs)\cite{san96}. High resolution near-IR and optical imaging of powerful AGN at low redshifts have revealed that their host galaxies are generally characterised by interactions (Fig.~\ref{fig_pgmorph}), with on-going or recent mergers, indicating that strong non-axisymmetric potentials are needed to induce the
high accretion rates\cite{vei06,vei09,kar10b}.

\begin{figure}[!h]
\centerline{\psfig{file=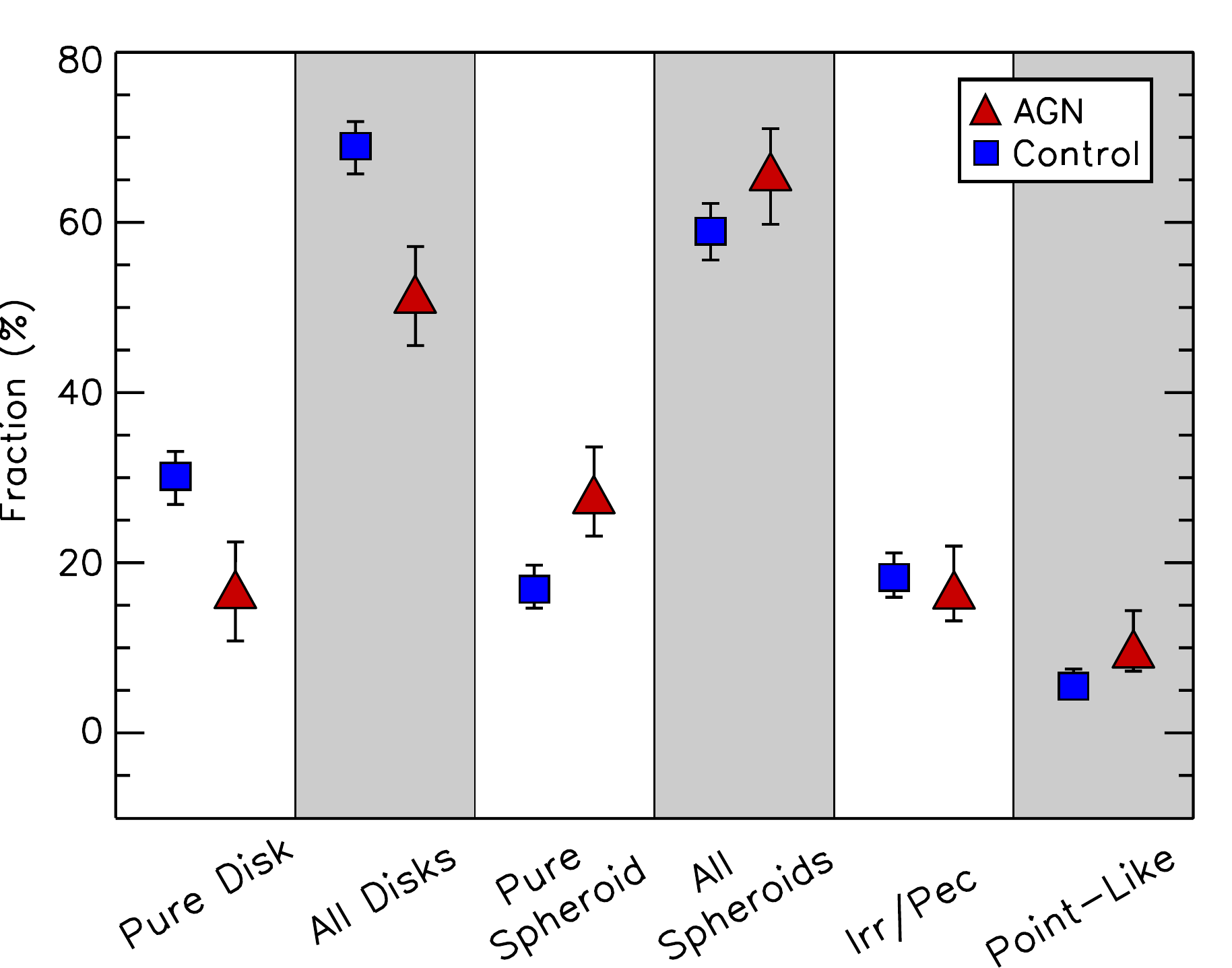,width=9cm}}
\vspace*{8pt}
\caption{Percentage of morphological classes for AGN and inactive galaxies (non-AGN), matched to have the same stellar mass distributions, at $z\sim2$. Figure adapted from Kocevski et al.~(2012)\cite{koc12}. Courtesy of Dale Kocevski. \copyright AAS. Reproduced with permission.  \label{fig_permorph}}
\end{figure}

The analysis of AGN host galaxy morphologies has now been extended to higher redshifts ($z\gsim1$),
especially thanks to the advent of the {\em Hubble Space Telescope (HST)} Wide Field Camera 3 (WFC3), which has enabled sensitive, high-resolution near-IR observations of significant samples of high-$z$ AGN. The general, and perhaps surprising result is that the host galaxies of most AGNs are in general not characterised by interactions or mergers: most of the AGNs are hosted in disks or bulge-dominated systems\cite{pie07,mai11,cis11,koc12,vil14}. This is the case even at $z\sim2$, when the global cosmic star formation and nuclear activity were at their maximum.  Figure~\ref{fig_permorph} shows the result of a morphological comparison study between 72 moderate-luminosity AGN and 216 non-active galaxies matched in stellar mass, all at $z\sim2$, from the {\em HST} CANDELS survey\cite{koc12}. The percentage of AGN and non-AGN in any of the morphological classes is very similar, and dominated by regular morphologies in all cases. The implications of these findings will be discussed in Section \S\ref{sec_agnsfr}.

\section{The AGN/Star-Formation Connection}
\label{sec_agnsfr}

Unveiling the co-existence of star formation and nuclear activity in galaxies through cosmic time, and disentangling their relative contributions to the galaxy IR emission, are key problems in Extragalactic Astronomy. The observed local relation between black-hole masses and galaxy bulge masses\cite{mag98,tre02,vol03} suggests that star formation and black hole growth must have been related at some point in the past, although this idea has been under scrutiny in recent times\cite{kor13}. Therefore, considerable efforts are being devoted to search for evidence of the co-existence of the two phenomena in nearby and distant galaxies.

\begin{figure}[h]
\centerline{\psfig{file=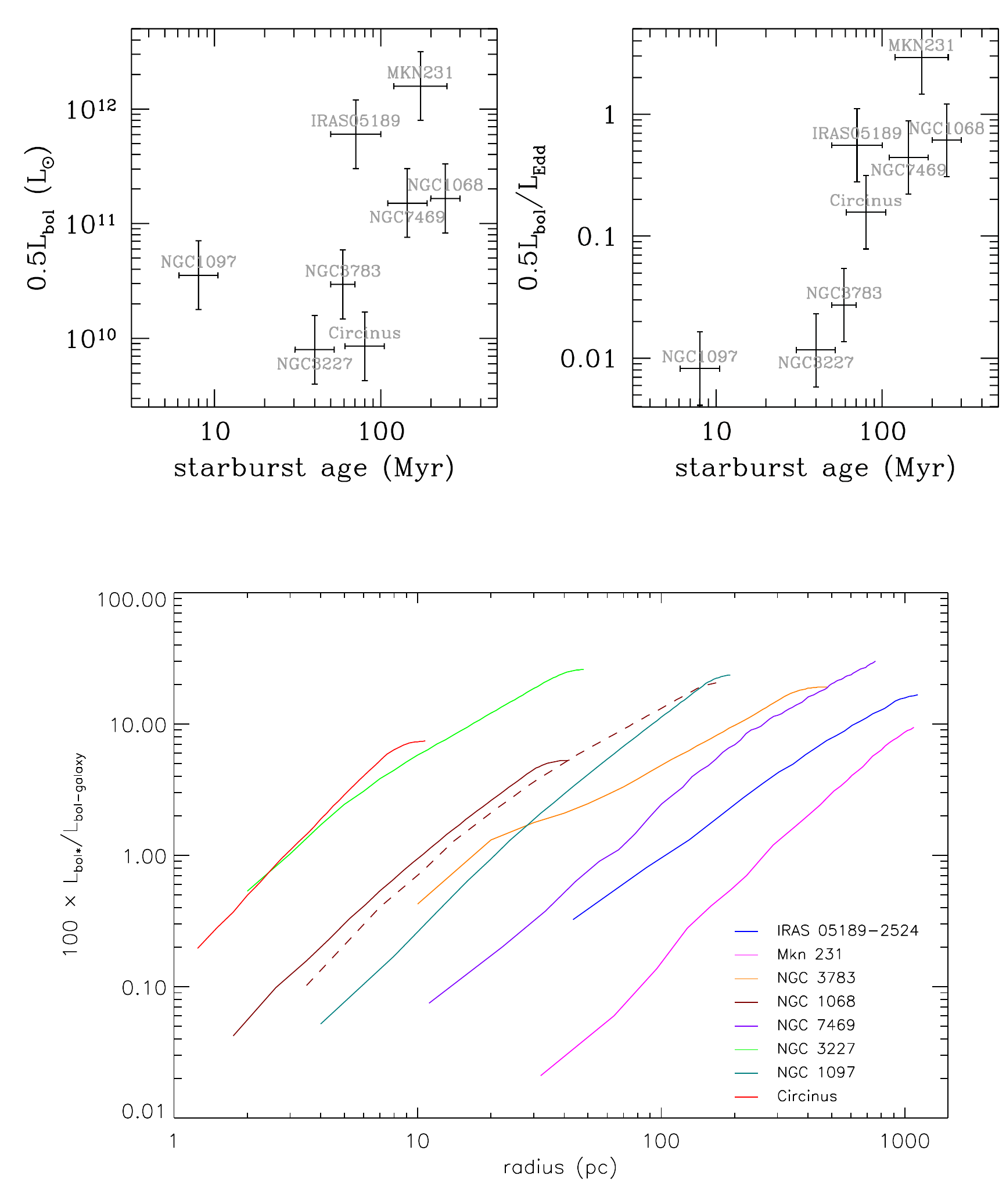,width=10cm}}
\vspace*{8pt}
\caption{{\em Top:} AGN luminosity, approximated as $0.5 \, \rm L_{bol}$,  versus the estimated age of the latest star formation episode in local Seyfert galaxies. {\em Middle:} the same plot, but for the relative luminosity with respect to the black-hole Eddington luminosity. {\em Bottom:}  The radial distribution of the relative bolometric luminosity contribution of young stars in these galaxies. Figures taken from Davies et al.~(2007)\cite{dav07}. \copyright AAS. Reproduced with permission.  \label{fig_davies} }
\end{figure}

The emission in different parts of the IR spectrum, from $\sim 1 \, \rm \mu m$ through  $~ 1 \, \rm mm$,  depends on the  dust temperature.  The dust heated by newly formed stars is characterised by low temperatures, typically $T \lsim 100 \, \rm K$, and its emission dominates the galaxy far-IR spectrum. The  AGN dusty torus can typically reach dust sublimation temperatures, i.e., $T\gsim 1000 \, \rm K$, and so its maximum emission occurs at near-/mid-IR wavelengths. Therefore, weighing the hot and cold dust components in a galaxy IR spectrum gives a good indication of the relative importance of star formation and black-hole activity. However, this decomposition is not trivial. The near-/mid-IR emission produced by the AGN overlaps with the direct emission from old stars in the host galaxy, and also the PAH emission, in case of on-going star formation.

\begin{figure}[h]
\centerline{\psfig{file=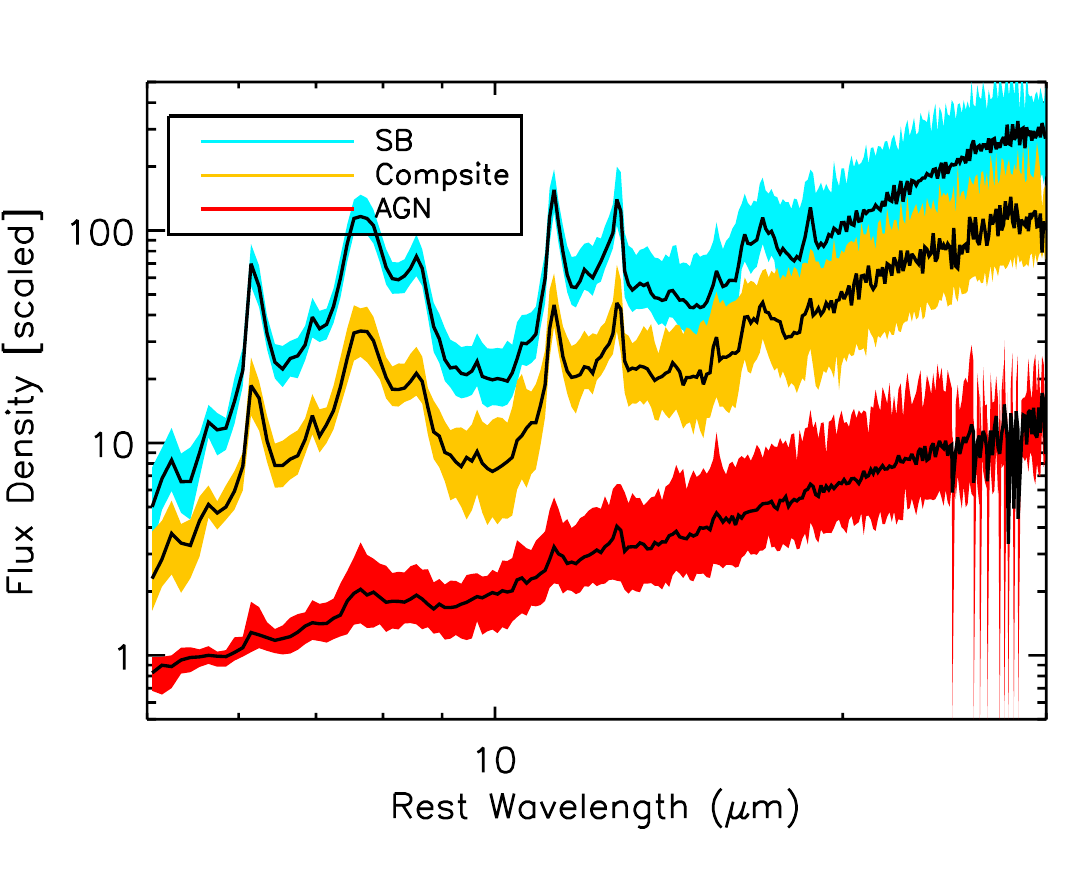,width=8cm}}
\vspace*{8pt}
\caption{Median {\em Spitzer} mid-IR spectra for local starbursts and AGN in the 5~mJy Unbiased {\em Spitzer} Extragalactic Survey. Figure taken from Wu et al.~(2010)\cite{wu10}. \copyright AAS. Reproduced with permission.  \label{fig_irssp} }
\end{figure}

\subsection{Follow up of known AGN at IR and sub-millimetre wavelengths}

Many observational studies have followed up known AGN in different parts of the IR spectrum, from near-IR through millimetre wavelengths, in order to investigate the presence and characteristics of star formation activity in their host galaxies.

For nearby galaxies, a major step in these studies has been produced by the use of adaptive optics (AO) on ground-based telescopes, which has enabled very high spatial-resolution observations at near-IR wavelengths.  Making use of the AO-assisted SINFONI spectrograph on the Very Large Telescope (VLT),  Davies et al. (Ref.~\refcite{dav07}) analysed the star formation activity of nine local Seyfert galaxies within very close scales to the central engine. They found  evidence for recent, rather than on-going,  star formation activity in these galaxies. The typical starburst age is of the order of $\sim 50-100 \, \rm Myr$ in almost all cases (Fig.~\ref{fig_davies}, top), which means that the black-hole fuelling lags the starburst onset in that amount of time. This result can be interpreted as the need for the starburst to slow down before significant accretion can be produced onto the central black hole: in a fresh starburst, the ejecta from supernovae and OB star winds travels too fast for this material to be accreted.

According to this same study, the stellar light profiles of Seyfert galaxies typically have sizes of a few tens of pc, and the stellar luminosity is only a minor fraction of the AGN luminosity within the inner 10 pc, but comparably important on kpc scales (Fig.~\ref{fig_davies}, bottom). Lower spatial-resolution studies conducted at mid-IR wavelengths  cannot disentangle such small scale effect, but rather detect the average star formation activity in scales of $\sim 0.1-1.0 \, \rm kpc$\cite{dia12,esq14}, except in very few cases\cite{mas12}.

At higher redshifts, the possibility of resolving the AGN circumnuclear region is quickly lost, but it is still possible to look for the main IR signatures of star formation activity in the host galaxies. Using the {\em Spitzer}\cite{wer04} Infrared Spectrograph (IRS)\cite{hou04}, several authors have searched for the presence of polycyclic aromatic hydrocarbon (PAH) emission at $3-20 \, \rm \mu m$. PAH emission is an unambiguous sign of star formation activity, so their detection in AGN hosts  directly argues for the simultaneous presence of the two phenomena.

On average, PAH equivalent widths are weaker in AGN hosts than in star-forming galaxies at similar redshifts\cite{wee05,bbr06,wu10} (see Fig.~\ref{fig_irssp}). However, PAH emission is still found in numerous AGN. From the spectroscopic study of a sample of 27 QSOs at $z<0.3$,  Schweitzer et al.~(Ref.~\refcite{sch06}) determined that these objects have similar PAH/far-IR and [NeII]/far-IR ratios to star-formation-dominated galaxies, and thus concluded that at least 30\% of their IR emission is due to star formation. This conclusion is consistent with the results of Shipley et al. (Ref.~\refcite{spl13}), who found that star formation accounts for up to 50\% of the total IR luminosity in $0.02<z<0.6$ IR galaxies that are AGN-dominated in the mid-IR. For other composite AGN/star-forming systems, where the AGN is present, but less dominant at mid-IR wavelengths, the contribution of star formation to the total IR luminosity is even higher\cite{nar08,alo12}.  For radio-selected AGN, the situation is less clear, as few mid-IR spectroscopic studies are available. Some powerful radio galaxies have clear PAH signatures in the IR spectra, while others do not\cite{sey08,raw13}. In any case, it is difficult to make a direct connection between the presence of an active nucleus and the PAH emission line intensities. In some galaxies most of this emission could be originated far from the nuclear region, and thus it would be hardly indicative of how PAH molecules behave close to the central engine.

\begin{figure}[h]
\centerline{\psfig{file=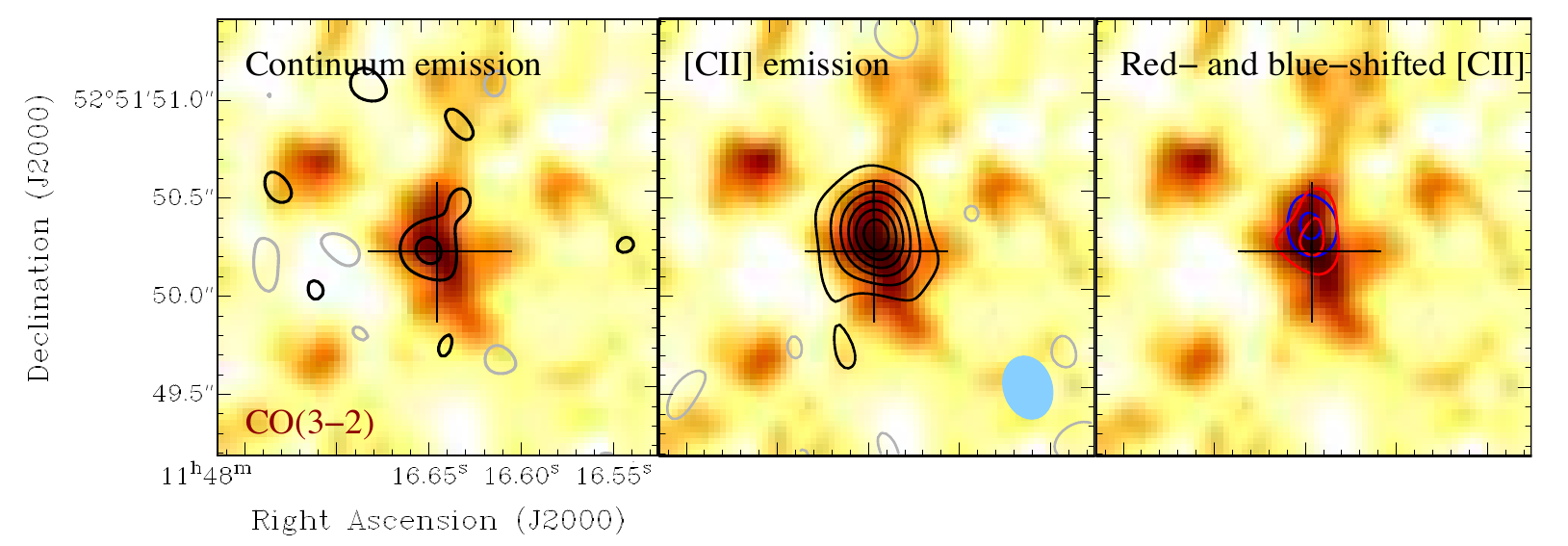,width=12cm}}
\vspace*{8pt}
\caption{Dust IR continuum and [CII] images of the quasar J1148+5251 at $z=6.42$. The spatial resolution is of $0.3 \times 0.2 \, \rm arcsec^2$, which corresponds to $1.7 \times 1.3 \, \rm kpc^2$  at that redshift.  Figure taken from Walter et al.~(2009)\cite{wal09}. Reprinted by permission from Macmillan Publishers Ltd: Nature, copyright (2009). \label{fig_walqso}}
\end{figure}

The follow up of known AGN at sub-millimetre wavelengths offers an alternative way of probing coeval star formation activity. Many sub-millimetre studies of powerful QSOs and other lower-luminosity AGN  at $z>1$ have been carried out over the last ten years\cite{vma05,stv05,bee06,wil07,cop08,mrs09,vig09}. A few pioneering works\cite{ber03a,ber03b,mai05} have even pushed these studies to rare QSOs at very high redshifts ($z \sim 6$), and now they are becoming increasingly common\cite{crl07,mai07,wan07,wal09,wan11,wan11b,omo13}. Walter et al. (Ref.~\refcite{wal09}) presented spatially resolved [CII] images of an IR hyper-luminous,  quasar host galaxy at $z=6.42$. They derived that the star-forming gas is distributed over a scale of $\sim 750 \, \rm pc$ (Fig.~\ref{fig_walqso}), which is around ten times larger than the typical starburst scale in local ULIRGs, such as Arp~220. This implies that this quasar host has a star-formation-rate (SFR) surface density comparable to the peak value in Arp~220 ($\sim 1000 \, \rm M_\odot \,  yr^{-1} \, kpc^{-2}$), but extended over an area 100 times larger. Such a high SFR implies that this system is forming stars at an Eddington-limit rate\cite{tho05}, likely forming a massive stellar bulge simultaneously with the massive black-hole fuelling. More recent studies have found other examples of this phenomenon in other rare luminous QSOs at $z\sim6$\cite{wan13}.

The observational evidence for the presence of large amounts of dust in QSOs (and galaxies) at $z\gsim 6$  is in principle puzzling, and imposes  important constraints on galaxy formation models. At low redshifts, most of the dust in galaxies is produced in the dusty envelopes of asymptotic giant branch (AGB) stars, but these stars typically need several hundred million years to be produced, so their formation by $z\sim6$ is not obvious. On the other hand, supernovae produce dust masses which are only $\sim 1\%$ of their progenitor star masses\cite{dwe11}, so they can unlikely be the only source of dust in high $z$ galaxies.  Recent detailed analyses of this problem\cite{dwe11,valr09} have shown that AGB stars could actually be responsible for a significant fraction of this dust production, but only if the galaxy had a very intense star formation activity (with $SFR \approx 2000 \, \rm  M_\odot/yr$), which should have dropped quickly  after the dust was produced.  In the case of AGN, though,  non-stellar mechanisms can be invoked for the dust production. Elvis et al. (Ref.~\refcite{elv02}) proposed that outflowing winds from the expansion of quasar broad emission line clouds can create dust, i.e., the observed dust is a consequence of the quasar activity, and is not necessarily linked to any powerful star formation, hence bypassing the problem of stellar evolutionary timescales for dust production (although in this scenario metals have to be available in the first place). This possibility was later considered by other authors\cite{mai06,mar07,pip11}.

\subsection{Weighing the AGN and star formation components in IR galaxies}

The most common approach to probe the simultaneous AGN/star-formation activity in large samples of IR galaxies is through the  modelling of their SEDs. Before the advent of the {\em Herschel Space Observatory}\cite{pil10}, the availability of far-IR photometry was mostly limited to low-$z$ galaxies, except for the bright and rare IR QSOs, and sub-millimetre galaxies known at that time. Thus, the first attempts consisted simply in adjusting a linear combination of known IR galaxy templates to the optical through mid-IR SEDs of AGN at different redshift.  These preliminary works have determined that a significant fraction of X-ray selected AGN show IR emission dominated by star formation in their host galaxies\cite{fra05,pol06}. Even when the presence of an AGN is evident from the SED mid-IR flux density excess, a significant part of the mid-IR light can be explained by star formation\cite{mur09}.

In the {\em Herschel} era, studies of the AGN/star-formation connection based on IR SED analysis have become more common. Hatziminaoglou et al. (Ref.~\refcite{hat10}) studied 79 AGN with robust detection in some of the {\em Herschel} far-IR bands, and concluded that a star-formation component is {\em always} needed to reproduce the entire IR SED of these sources. The IR luminosities associated with star formation and accretion in $z>2$ objects appear to have a $\sim$90\% correlation probability, following $L_{\rm SB} \propto L_{\rm accr}^{0.35}$.

Other {\em Herschel}-based studies have investigated the presence and importance of AGN among the more general population of IR-selected galaxies. Special attention has been paid to the study of galaxies at $z\sim 2$, as this is, on average, the main epoch of star formation and nuclear activity of the Universe.  The exact percentage of AGN incidence found among IR galaxies at $z\sim2$ varies according to different selection effects, particularly IR luminosities and IR colours, ranging from at least 20-30\% in the global ULIRG population\cite{gru10,poz12,saj12} to 50-60\% among the reddest galaxies\cite{mel12}. However, all studies agree that {\em the total IR emission of galaxies is dominated by star formation rather than nuclear activity}.

A critical point of discussion over the last few years has been whether AGN host galaxies have enhanced, suppressed, or equally important star formation activity, with respect to galaxies with no nuclear activity. Several authors have tackled this problem, but inferred quite different conclusions, mainly due to sample variance effects. A recent study of a small sample of luminous quasars (with X-ray luminosities $L_X>10^{44} \, \rm erg \, s^{-1}$) in the {\em Herschel} bands has claimed that star formation was suppressed in these sources\cite{pag12}. This result is at odds with other works that found that powerful radio sources do host significant star-formation activity\cite{bar12,har12}. Harrison et al. (Ref.~\refcite{har12}) derived obscured SFR for AGN selected in three different fields of the sky at $1<z<3$, versus their X-ray luminosities. They found that the mean SFR of AGN hosts are comparable to those of galaxies with no nuclear activity in the upper envelope of the SFR/stellar mass relation (SFR-M$^*$), which is sometimes called the  `main-sequence' of star formation.  However, the scatter was sufficient to explain the apparently discrepant results found by other AGN/star-formation studies based on lower statistics.

\begin{figure}[h]
\centerline{\psfig{file=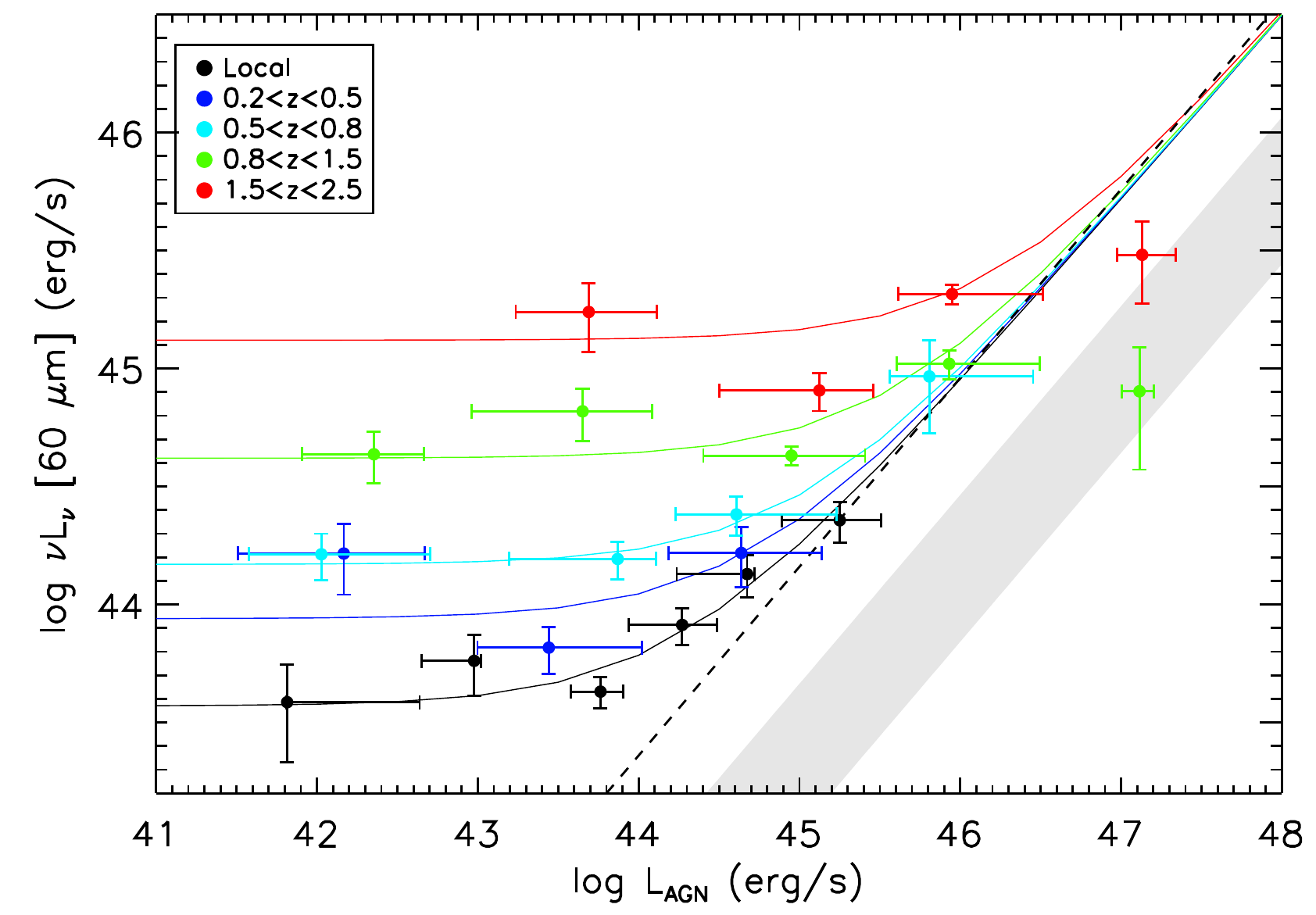,width=12cm}}
\vspace*{8pt}
\caption{Mean rest-frame $60 \, \rm \mu m$ luminosities versus X-ray luminosities for AGN at different redshifts. The rest-frame $60 \, \rm \mu m$ luminosities can be considered proportional to the host galaxy SFRs, as the AGN emission has a very minor contribution at these wavelengths.   It is evident that $\nu L_\nu(60 \, \rm \mu m)$ increases with X-ray luminosity only for the most X-ray luminous AGN up to $z\sim1$, while no trend is observed in the other cases.  Figure taken from Rosario et al.~(2012)\cite{ros12}. Reproduced with permission from Astronomy \& Astrophysics, \copyright ESO. \label{fig_rossfr} }
\end{figure}

More general comparisons of the {\em Herschel}-based SFR of X-ray-selected AGN and IR  galaxies with no nuclear activity have shown that the most luminous X-ray AGN (with $L_X \gsim 10^{43.5} \, \rm erg \, s^{-1}$)  have a significant enhancement of their star formation activity, while less luminous AGN do not\cite{sha10,lut10,san12}. In a more extended analysis, Rosario et al. (Ref.~\refcite{ros12})  found that, actually, the star-formation enhancement for high X-ray luminosities is observed up to $z\sim1$, while at higher redshifts no significant trend is observed (Fig.~\ref{fig_rossfr}).  This is consistent with other {\em Herschel}-based results that claim that active and inactive galaxy hosts at $z\sim2$ are statistically indistinguishable\cite{mul12,ros13b}, and with the morphological results discussed in Section \S\ref{sec_hosts}.

These results have been interpreted as evidence for a change in the main driver for  AGN fuelling with luminosity and redshift: at $z<1$, major mergers may  be responsible for a synchronised black-hole and stellar-bulge growth in luminous AGN, while low-luminosity AGN simply grow through `stochastic fuelling' (i.e., any non-major-merger processes)\cite{hh06}.  At $z>1$,  if this division still exists, it must  have moved to higher luminosities than those typically observed in AGN surveys. This can be explained by the fact that gas reservoirs were larger, and thus massive black-hole fuelling did not require any particular non-secular process.

Radio-selected AGN likely correspond to a different population than X-ray AGN, and presumably are in a later evolutionary stage.
Del Moro et al. (Ref.~\refcite{delm13}) studied the star formation properties of radio-excess selected AGN, and found that their specific SFR (i.e. the SFR divided by the respective stellar masses), are lower than for X-ray-selected AGN. These results are in line with other radio galaxy studies\cite{dic12}, although other authors have found no statistical difference in the far-IR properties of radio-loud and radio-quiet galaxies\cite{har10}.

Overall, this wide variety of AGN follow up results at IR wavelengths indicate that the co-existence of star formation and nuclear activity in galaxies is a common phenomenon. However, it is still difficult to draw a firm conclusion on  whether, or when, the two phenomena were actually `synchronised' at different redshifts.  Put another way, the co-existence of the two phenomena does not necessarily imply causality. The lack of an enhancement in the star formation activity of typical AGN at $z>1$, and their predominantly regular morphologies, including a high disk incidence, suggest that these galaxies did not suffer a recent major merger that leaded to a common black-hole/stellar bulge growth. Disk instabilities at high redshifts could have been sufficient to keep typically luminous AGN feeding\cite{boud11}. Note, however, that the importance of mergers cannot be excluded,  as the relaxation times could be faster at high redshifts than at $z<1$. It is then plausible that black hole growth is related to host galaxy growth in some cases, but this may not be a universal fact.

\section{AGN feedback}
\label{sec_feed}

Feedback is the fundamental process that self-regulates star formation in galaxies, and is a key ingredient in galaxy formation models. Positive feedback favours further star formation activity. Instead, negative feedback prevents galaxies from over-growing. Feedback from star formation processes (supernovae, radiation pressure, stellar winds and photoionisation) is expected to dominate in galaxies with low stellar masses ($\rm M<M^*$), while AGN are thought to be the main feedback drivers in massive galaxies. More specifically, during the very luminous AGN phases ({\em radiative mode}) the radiation pressure is expected to drive a powerful, very energetic, nuclear wind, which shocks and accelerates the interstellar medium (ISM) of the host galaxy, producing massive outflows that remove a substantial fraction of the available gas and quench star formation. Direct radiation pressure on the dusty clouds can also be an alternative, or additional, way to drive outflows in AGN host galaxies, contributing to star formation quenching. Subsequently, in the {\em radio mode}, injection of energy into the galaxy halo by the radio jets prevents the halo gas from cooling and replenish the galaxy with fresh fuel for additional star formation. According to models, this `maintenance mode' is responsible for keeping massive galaxies clean, and making them evolve passively into today's red-and-dead massive ellipticals\cite{hic09}. Other models also invoke occasional (close-to-Eddington) accretion onto the black hole, in short quasar-mode bursts, resulting from accretion of gas simply released by the stellar winds in the host galaxy, which further help in keeping the galaxy ISM clean of gas and dust\cite{cio10}.

\begin{figure}[!h]
\centerline{
\includegraphics[scale=0.3]{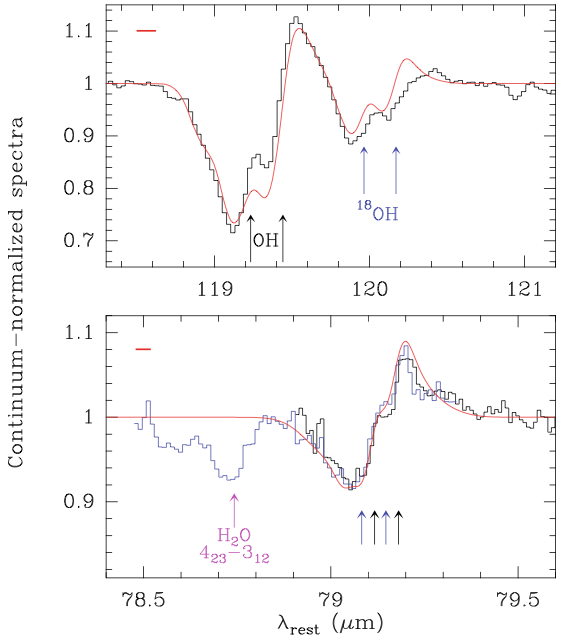}}
\caption{{\em Herschel}/PACS far-IR spectrum of Mrk231 showing a prominent P-Cygni profile in the OH and H$_2$O transitions, revealing a massive molecular wind with velocities of about 1000 km/s. Figure taken from Fischer et al.~(2010)\cite{fis10}. Reproduced with permission from Astronomy \& Astrophysics, \copyright ESO.\label{fig_fisch}}
\end{figure}

\begin{figure}[!h]
\includegraphics[scale=0.25]{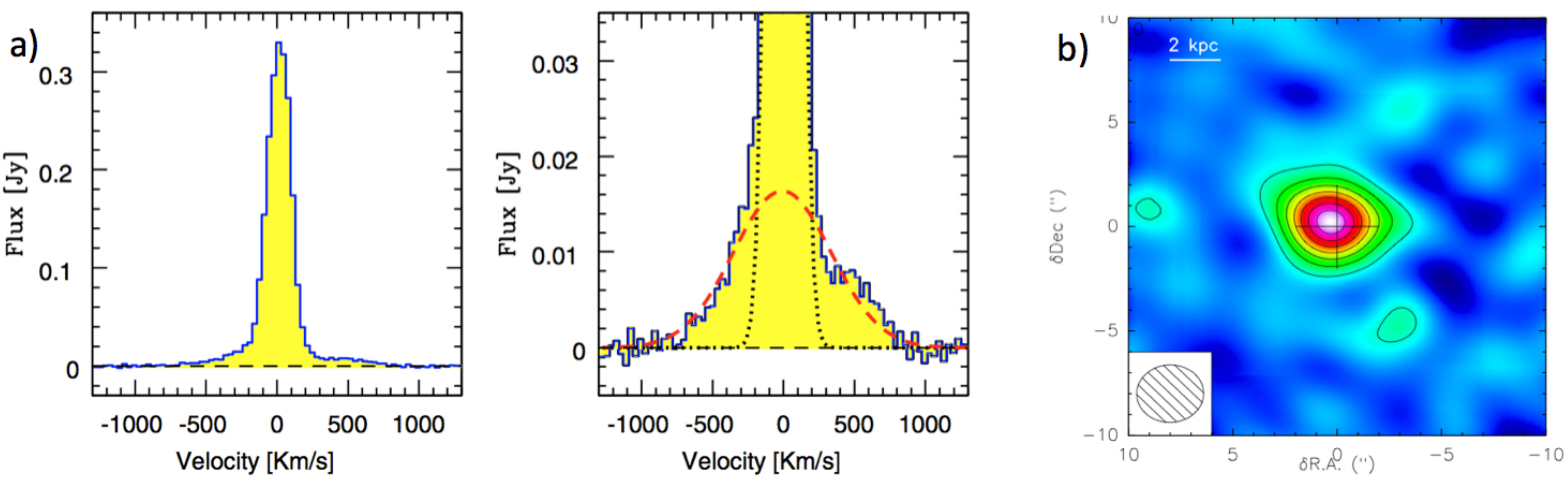}
\caption{a) Plateau de Bure Interferometer spectrum of the CO(1--0) transition in Mrk231 revealing very broad wings, tracing the same outflow observed in absorption in the OH far-IR transitions (see Fig.~\ref{fig_fisch}). b) The map of the CO(1--0) broad wings reveals that the outflow is extended on kpc scales. Figures taken from Feruglio et al.~(2010)\cite{fer10}. Reproduced with permission from Astronomy \& Astrophysics, \copyright ESO.  Panel b) is courtesy of Roberto Maiolino. \label{fig_ferug}}
\end{figure}

Observational evidence for AGN feedback in the radiative mode, consisting of powerful massive outflows
driven by AGN, has only been found recently.$\,${\em Herschel}  has discovered spectacular
P-Cygni profiles of some far-IR molecular transitions, such as OH at 79$\mu$m, in local galaxies hosting
powerful AGN\cite{fis10,stu11,spo13,vei13} (Fig.~\ref{fig_fisch}). Such P-Cygni
profiles, with blueshifted absorption velocities in excess of 1000~km/s are a clear signature of massive
molecular outflows. The signature of the same molecular outflows (often in the same objects)
has also been detected through broad emissions wings of CO millimetric transitions\cite{fer10,cic12,com13}, using sub-/millimetre interferometric observations (Fig.~\ref{fig_ferug}). These observations have revealed that such outflows are extended on kpc scales. The inferred outflow rates are very large, approaching $1000 \, \rm M_{\odot}~yr^{-1}$ in some cases.
The kinetic power of these winds is of the order of a few percent of the AGN bolometric
luminosity, and the momentum rate is of the order of $\rm 20 \, L_{AGN}/c$, in very nice agreement with
the expectations of quasar driven winds. Observations of molecular transitions tracing the dense phase of the molecular gas, such as HCN, have revealed that such quasar driven winds do host large amounts of dense gas\cite{aal12}.

These observations in nearby quasars provide an excellent laboratory to investigate quasar feedback.
However, if this mechanism has to explain the properties of local red-and-dead elliptical galaxies, then
the bulk of the action has to happen at high redshifts, at which  IR observations have also found evidence for quasar feedback. Maps of nebular lines redshifted into the near-IR have revealed powerful outflows extended
on scales of several kpc\cite{ale10,cano12}. More importantly, observations of the [CII]$158 \, \rm \mu m$ far-IR fine structure line in some of the most distant quasars (z$>$6) have revealed outflows of atomic gas with outflow rates exceeding $3000 \, \rm M_{\odot}~yr^{-1}$, and extending on scales of $>$10~kpc\cite{mai12}, implying that quasar feedback was already in place in the early Universe. This could explain the presence of massive and passive galaxies already present by redshifts z$\sim$2-3.

In the radio mode, the galaxy molecular gas is expected to be heated by energy injected by the radio jets. Sub-/millimetre  observations have shown that this is the case for radio galaxies, which have substantial molecular gas reservoirs\cite{das12}, but not always high SFR\cite{nes11}. In fact, a fraction of the gas in the halo can cool down, forming molecular gas that streams towards the  galaxy\cite{sal11}, although the inflow rate is probably modest. Both radio and millimetre observations are also revealing that radio jets not only heat gas in the halo, but also affect the ISM of the host galaxy by ejecting atomic and molecular gas\cite{mor13}.

\section{Statistical Studies of AGN through Cosmic Time}
\label{sec_stat}

\subsection{IR techniques to identify AGN in large galaxy samples}
\label{agnsel-midir}

\subsubsection{IR colour-colour techniques and SED analysis}

The simplest and most widely-used methods to select AGN within large IR galaxy samples are mid-IR colour-colour diagrams. The colour segregation is produced by the power-law spectral shape characteristic of the AGN hot, dusty-torus emission, which in some cases dominates the spectral energy distribution (SED) of the host galaxy at mid-IR wavelengths. The most commonly adopted versions of these colour-colour diagnostics\cite{lac04,ste05}  have been calibrated using some of the earliest blank galaxy surveys conducted with the {\em Spitzer} Infrared Array Camera (IRAC)\cite{faz04} . These criteria resulted particularly reliable to identify luminous type-1 AGN, but less effective to identify type-2 AGN, whose colours are more similar to those of many normal galaxies. More recently, updated versions of the IRAC colour-colour criteria improved the selection of luminous type-2 AGN\cite{lac07,don12}.  At the same time, other works have adapted these colour criteria to the {\em WISE} telescope mid-IR filters\cite{mts12,ste12,assef13}.  All these colour-colour selection techniques have been calibrated with relatively shallow mid-IR and X-ray data. Consequently, they are optimised to select luminous AGN at different redshifts, but they dramatically lose completeness at X-ray luminosities $L_X \lsim 10^{43} - 10^{44} \, \rm erg \, s^{-1}$, and/or when the light of the host galaxy also makes a significant contribution to the mid-IR SED\cite{bar06,car08,bru10,cap13}.

\begin{figure}[pb]
\centerline{\psfig{file=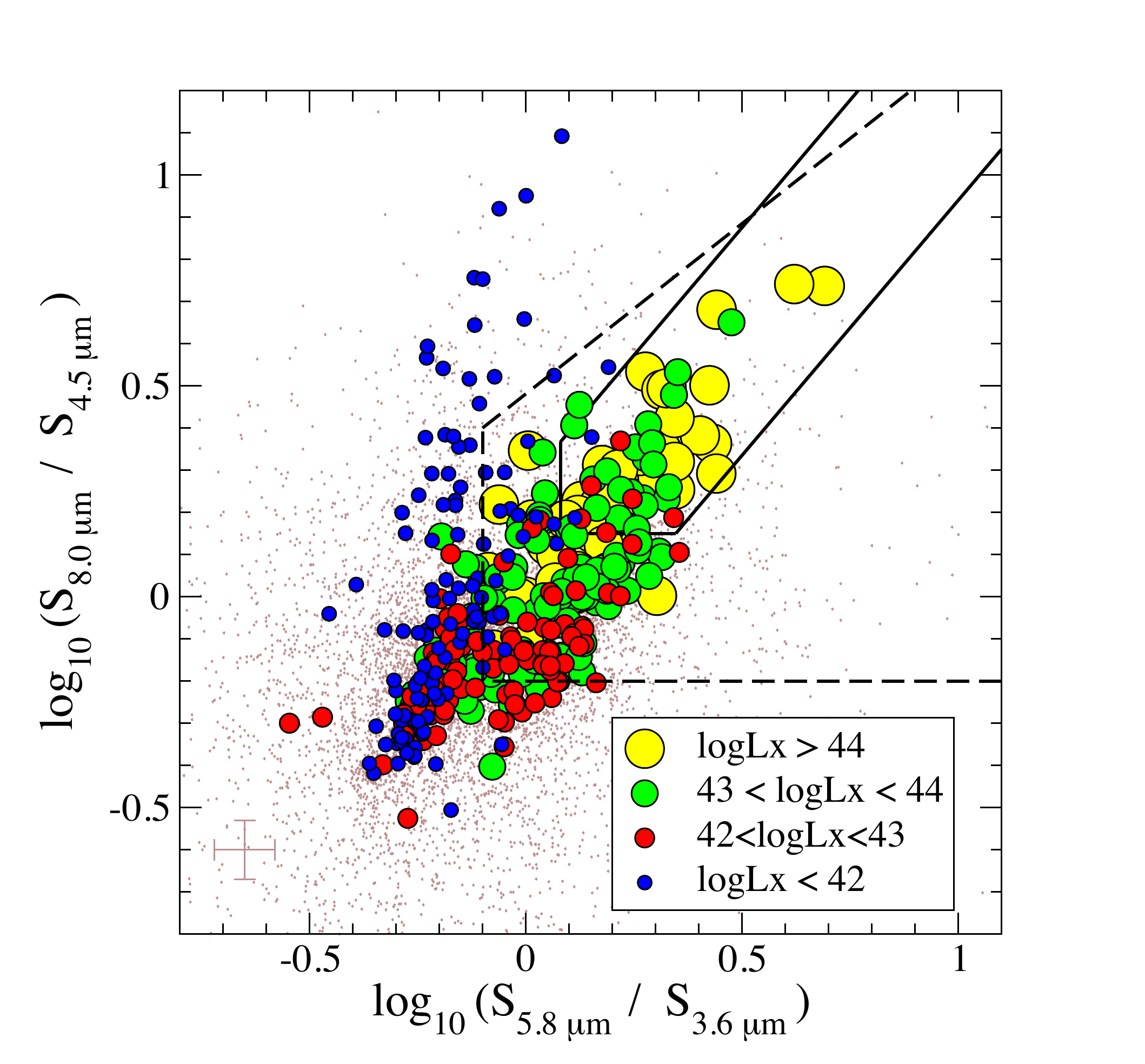,width=11cm}}
\vspace*{8pt}
\caption{IRAC colour-colour diagram showing the location of X-ray AGN with different X-ray luminosities in the Chandra Deep Field South 4Ms catalogue\cite{xue11} (circles), with respect to other IRAC galaxies not detected in X-ray, or detected but not classified as AGN (background dots). The dashed and solid lines delimit the colour regions proposed for AGN selection by Lacy et al.\cite{lac07} and Donley et al.\cite{don12}, respectively. \label{fig_iraccol}}
\end{figure}

Figure~\ref{fig_iraccol} shows the IRAC colours of the X-ray AGN in the Chandra Deep Field South 4Ms catalogue\cite{xue11}. It is evident how the colours vary with X-ray luminosity: the more luminous X-ray AGN display, on average, significant redder IRAC colours than less luminous ones.  The vast majority of AGN with $L_X < 10^{44} \, \rm erg \, s^{-1}$ occupy the same locus as normal galaxies in the IRAC colour-colour diagram, so the strictest colour criteria are unable to select moderate luminous AGN, while less strict colour selections inevitably have a large fraction of contaminants.

The AGN selection through IR colour-colour techniques has recently been extended to the far-IR regime using {\em Herschel} data. Far-IR data at $\lambda> 300 \, \rm \mu m$ alone appears not to be adequate for such a purpose\cite{hat10}. A combination of mid- and far-IR data is more effective instead\cite{fre04,kir13}.  Nonetheless, the general effectiveness of these mid-/far-IR colour-colour diagrams for low-luminosity AGN and highly obscured AGN, remains to be proved.

An alternative, more reliable method to identify AGN among IR galaxies is directly modelling their SEDs. Alonso Herrero et al. (Ref. \refcite{alo06}) studied a sample of 97 $\rm 24 \, \rm \mu m$-selected galaxies which display a power-law shape in their IRAC SEDs. Within their sample,  half of the pure power-law sources are X-ray confirmed AGN, and the other half are not, but many are likely IR-bright, X-ray obscured AGN (the X-ray data used for this identification was shallower than that used in Fig.~\ref{fig_iraccol}). This power-law SED technique has also been applied by several other authors to investigate the presence of AGN among IR galaxies\cite{bar06,car08,kar10}. More recently, this SED analysis technique has been generalised to identify the presence of AGN not only in pure power-law galaxies, but also in composite systems, where the power law is a significant component of the mid-IR SED, but not necessarily dominant to the host galaxy\cite{cap13}.  The power-law IR SED analysis results in an efficient way of identifying AGN in large galaxy samples, especially when no X-ray data are available.

Finally, another common approach consists in trying to fit the galaxy SEDs with local  AGN templates (either in pure form, or in linear combination with stellar templates). Several works have explicitly produced updated galaxy template libraries that include AGN and/or galaxy/AGN composite systems \cite{pol06,sal09,assef10,lee13,hao14}.

\subsubsection{IR spectral diagnostics}

{\em Spitzer} mid-infrared  spectroscopy has also been very important to study the presence of AGN in IR sources.
However, in contrast to the widely applicable photometric techniques,  spectroscopic studies could so far only be conducted over relatively small and bright galaxy samples. I discuss here the importance of different IR spectral features as AGN indicators, and recently-proposed IR spectral diagnostics for AGN.

There is a series of broad mid-IR spectral features, which combined, can provide a good indication of the presence of an active nucleus. These are basically  the lack or weakness of PAH emission lines, the power-law continuum shape, and in some cases a strong silicate absorption. Low resolution spectroscopy ($R\sim500$) is sufficient to study these features, and thus this kind of diagnostic has been widely adopted in the analysis of {\em Spitzer} spectroscopic data  \cite{dal06,saj07,nar10,pet11}. The presence of all these  mid-IR spectral features can be explained with radiative transfer models\cite{hey12,sny13}. 
 
 Another point of discussion in the literature has been whether the PAH emission line ratios are different in galaxies hosting nuclear activity. From theoretical considerations, the $L(7.7 \, \rm \mu m)/L(11.3 \, \rm \mu m)$ ratio is expected to vary with the PAH ionisation balance\cite{dra07}, and thus expected to be lower in the vicinities of the central black hole. At low redshifts, some authors have found a smaller  $L(7.7 \, \rm \mu m)/L(11.3 \, \rm \mu m)$ and other PAH line ratios in AGN than in star forming galaxies\cite{smi07,odo09,wu10,mag13}. Instead, other authors found weak or no relation between these line ratios and the importance of nuclear activity \cite{sal10,lam12,shi13}.  Once more, the interpretation of these different results is hampered by the inability of resolving small spatial scales. As discussed before, PAH emission could be produced far away from the nuclear region, and thus not be indicative of the effect of the active black hole on PAH molecules.

Other studies have also incorporated high ionisation lines, which are more secure AGN indicators. For example, the presence of [OIV] at  $\lambda= 25.89 \, \rm \mu m$, and especially [NeV] at $\lambda=14.32 \, \rm \mu m$,  can be used alone as an AGN diagnostic, as the ionisation potentials of these transitions require a hard radiation field that almost exclusively AGN can produce (at least to be significant in the integrated galaxy spectrum). Moreover, the  [NeV]/[NeII] and [OIV]/[NeIII] ratios are commonly used as proxies of the AGN contribution to the total galaxy IR luminosity\cite{arm06,pet11,stu02,fah07,das08,tom10,wea10}.

\begin{figure}[h]
\centerline{\psfig{file=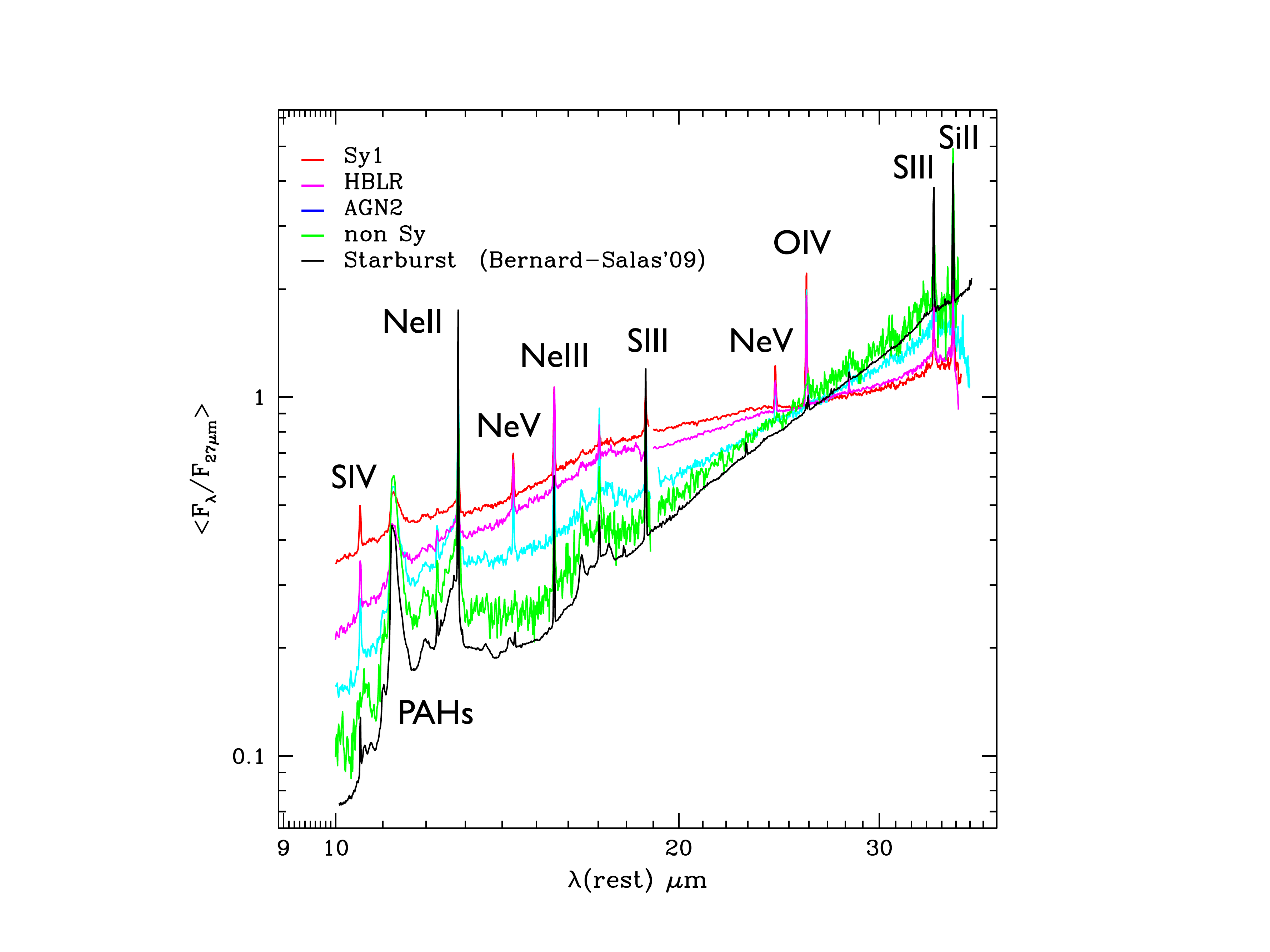,width=12cm}}
\vspace*{8pt}
\caption{Composite high-resolution, mid-IR spectra for different types of AGN\cite{tom10}, compared to the typical spectrum of normal star-forming galaxies\cite{ber09}. All main fine structure lines, and PAHs, are indicated. Figure adapted from Tommasin et al.~(2010)\cite{tom10}. \copyright AAS. Reproduced with permission.  \label{fig_agnspec}}
\end{figure}

Figure~\ref{fig_agnspec} shows composite high-resolution, mid-IR spectra for different types of AGN\cite{tom10}, compared to the typical spectrum of non-Seyfert galaxies, and other normal star-forming galaxies\cite{ber09}. It is clear that high-ionisation lines like [NeV] and [OIV] are only prominent in AGN, while the equivalent widths of PAH features are typically small.

Observing these high ionisation lines in high redshift galaxies requires very sensitive instruments that can operate more into the far-IR regime ($\lambda \gsim 40 \, \rm \mu m$). Future far-IR telescopes, such as SPICA\cite{nak07} and Millimetron\cite{smi12}, should open the possibility of conducting systematic studies of highly obscured AGN at $z>1$\cite{fra10,spi12}. It should be noted that, especially at high redshifts, these IR diagnostics become crucial for an unbiased AGN identification in galaxy samples, as deeply embedded AGN could be elusive and remain undetected even in very deep X-ray surveys.

\subsubsection{Multi-wavelength diagnostics to unveil AGN among IR galaxies}

Multi-wavelength spectral diagnostics have  provided very important information on the presence of AGN among IR-selected galaxies. This is particularly the case of multi-object optical spectroscopy, which allows for the simultaneous observation of large galaxy samples.  For type-1 AGN, the characteristic broad lines produced by the gas clouds close to the central engine can be observed in their optical spectra\cite{pap06,lac07,cap08,ram13}. IR-selected type-2 AGN are usually determined by their location in the BPT  diagram\cite{bpt81}, and other line ratios, which indicate a powerful dust heating mechanism produced by central accretion\cite{kau03,kew06}. These spectral classifications have been applied  to mid-IR-selected galaxies in different fields of the sky\cite{cap08,cap09,lam10,sym10,jun11,pat11,cas12}. 

The joint analysis of IR and radio data is also a powerful tool to unveil the presence of AGN among IR sources. The ratio between the IR and radio 1.4~GHz luminosities (i.e. the $q_{\rm IR}$ parameter)\cite{vdk73,con82} is typically used to discriminate between galaxies dominated by nuclear activity and star formation, as radio-loud AGN are characterised by relatively low values of this ratio\cite{ibar08,par10}. For star-formation dominated galaxies, the $q_{\rm IR}$ parameter remains quite constant up to at least $z=1$, within a dispersion factor $\lsim 1.5$ (see Refs. \refcite{app04,jar10}), and it only shows a  tentative, mild decline\cite{ivi10a,ivi10b,huy10}, or basically none at higher redshifts \cite{sar10,bou11}.

In fact,  Sargent et al. (Ref. \refcite{sar10}) recently argued that both star-forming galaxies and (radio-quiet) AGN populate the IR-radio luminosity correlation, which suggests that the two phenomena are coevally present in many IR galaxies, and/or are both responsible for the observed trend. In addition, it has been suggested that AGN occupy two relatively distinct branches in the plane of mid-IR and radio luminosity. Seyfert galaxies lie almost exclusively on a mid-IR-bright branch, while low-ionisation nuclear emission line galaxies (LINERs) are split evenly into a bright and a faint branches\cite{ros13}.

Other authors have studied the connection between the infrared and hard ($>10 \, \rm keV$) X-ray properties of local AGN using the {\em Swift} and {\em AKARI} all-sky surveys\cite{ich12,mat12}. They found a good correlation between hard X-ray and mid-IR luminosities for both X-ray absorbed and unabsorbed AGN, which suggests the presence of clumpy dusty tori producing isotropic IR emission. 

\begin{figure}[h]
\centerline{\psfig{file=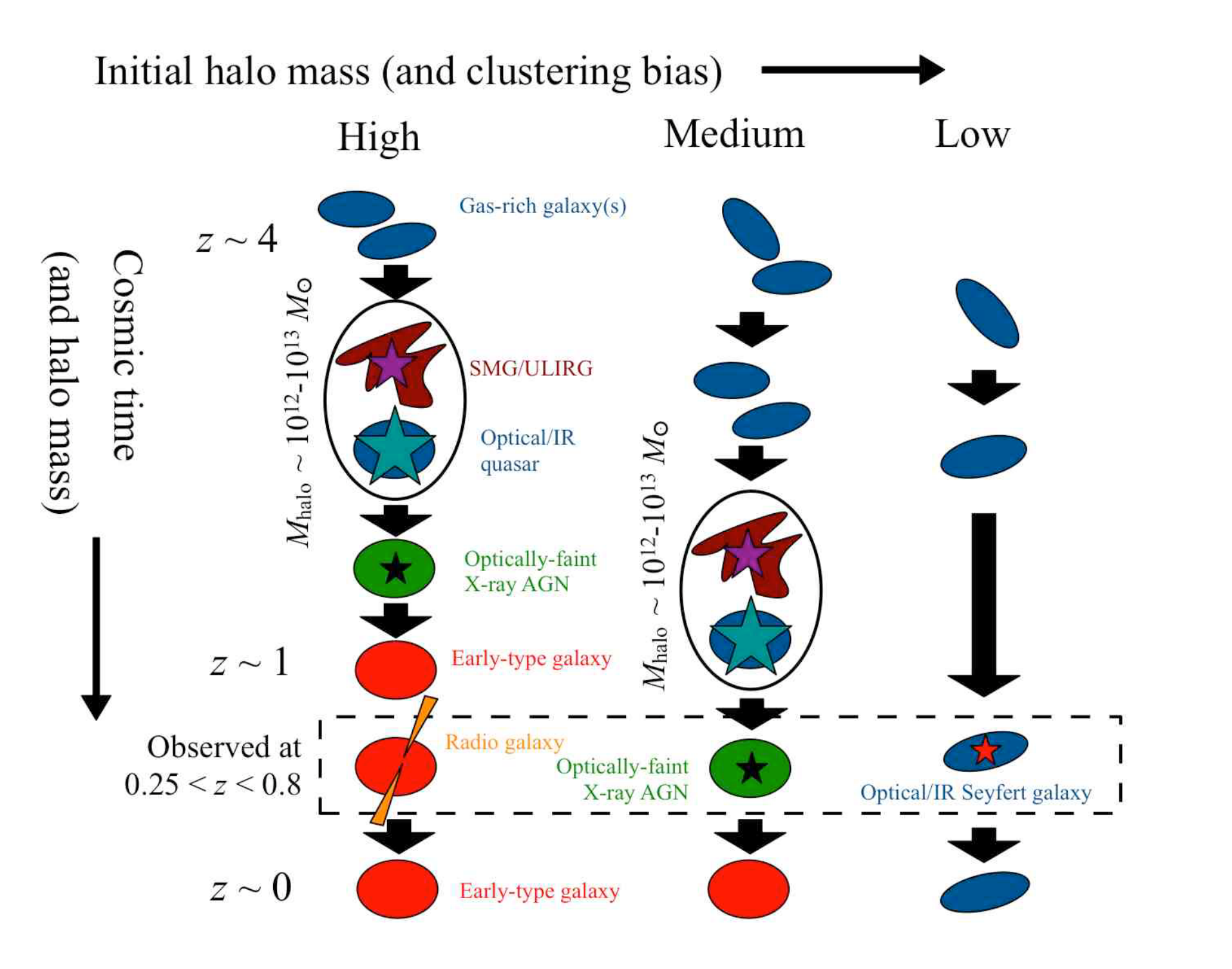,width=12cm}}
\vspace*{8pt}
\caption{ \label{fig_agnseq} Cartoon illustrating a possible sequence for AGN/host galaxy evolution, according to the initial dark matter halo masses. Figure taken from Hickox et al.~(2009)\cite{hic09}. \copyright AAS. Reproduced with permission. }
\end{figure}

Arguably a more complete approach consists in combining IR, radio and X-ray altogether for a more comprehensive study of AGN in large galaxy samples. By applying such a multi-wavelength analysis to $\sim 600$ AGN at $0.25<z<0.8$, Hickox et al. (Ref.\refcite{hic09}) attempted to reconstruct the AGN evolutionary sequence  from a radiatively efficient phase (radiative mode) to a radiatively inefficient phase (radio mode). This sequence has been previously proposed, or considered, in different theoretical studies\cite{cro06,mon07,hop08,som08}. Fig.~\ref{fig_agnseq} illustrates this possible evolution. The observed galaxies at $0.25<z<0.8$ display different degrees of clustering, which allows one to infer that they reside in dark matter haloes of different masses. The evolutionary path is proposed to vary with halo mass: in haloes with masses above a critical value ($\gsim 10^{12} \, \rm M_\odot$), luminous AGN are formed. They pass through IR and X-ray bright phases, which result in the formation of early-type galaxies (the more massive the halo, the earlier this evolution).  Instead, Seyfert galaxies are the product of secular evolution in low-mass haloes.  Note, however, that the galaxy merger scenario suggested by this cartoon should be taken with care, as the latest studies suggest a relatively minor role of galaxy mergers in fuelling nuclear activity at redshifts $z\sim1-3$ (see Section \S\ref{sec_hosts}). Therefore, confirmation of how these evolutionary paths proceed will require further observational constraints at $z>1$.

\subsection{AGN among IR galaxies through cosmic time}
\label{agnsel-evol}

\subsubsection{The Low-Redshift Universe ($z<1$)}

A wide range of studies have analysed the incidence and importance of AGN among IR galaxies at low redshifts, including the local Universe. From redshifts $z=1$ to $z=0$, the extragalactic IR background is increasingly dominated by normal IR galaxies (with total IR luminosities $L_{\rm IR}<10^{11} \, \rm L_\odot$), while luminous IR galaxies (LIRGs) and ULIRGs, with $10^{11} < L_{\rm IR}<10^{12} \, \rm L_\odot$ and $L_{\rm IR}>10^{12} \, \rm L_\odot$, respectively, decline rapidly in number density\cite{lef05,cap07}. Overall, this is the consequence of the global decline of the SFR density and powerful nuclear activity in the Universe. 

\begin{figure}[t]
\centerline{\psfig{file= 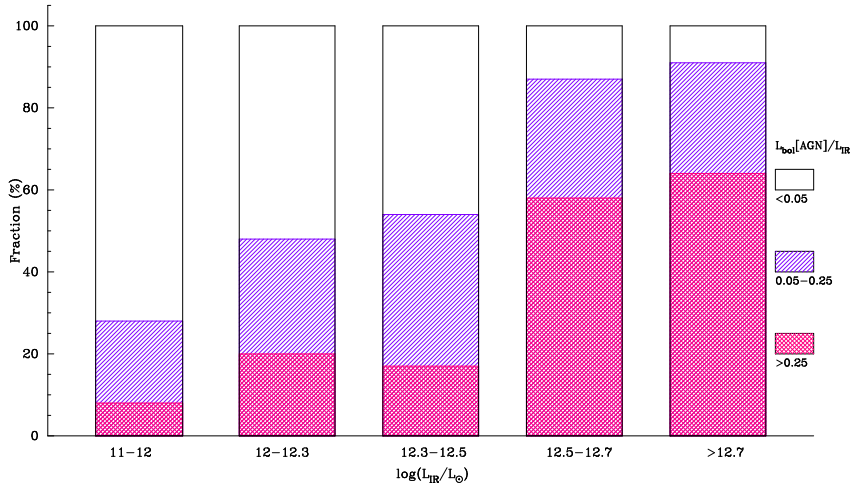, width=14cm}}
\vspace*{8pt}
\caption{Fraction of local IR sources with different AGN contributions to the total IR luminosities (x axis). These AGN contributions are colour-coded as shown on the right-hand side of the plot. Figure taken from Alonso-Herrero et al.~(2012)\cite{alo12}, including results from Nardini et al.~(2010)\cite{nar10}. \copyright AAS. Reproduced with permission.  \label{fig_agnlumhis}} 
\end{figure}

The incidence of AGN among local LIRGs and ULIRGs is quite important, as up to $50\%-60\%$ of them could contain an AGN\cite{ris10,pet11,alo12}. This result is consistent with the outcome of recent  studies of $70 \, \rm \mu m$-selected LIRGs at redshifts $0.3<z<1$\cite{jun13}. When considering also normal IR galaxies, the fraction of AGN among IR sources appears to be somewhat lower, i.e. around 30\%\cite{gou09}. These percentages should be taken with some caution, as they depend significantly on the IR luminosities of each galaxy sample, and  also on the criteria used to identify the AGN presence. Energetically, though, it is clear that AGN only make a small part of the total IR background budget at low redshifts:  only $\sim 10-15\%$ at $z\sim1$\cite{jau11,wu11}. At $z\sim0$, only the most luminous ULIRGs have, in general, more than 25\% of their bolometric IR luminosity dominated by an AGN component\cite{alo12,nar10} (Fig.~\ref{fig_agnlumhis}). Between $z=1$ and $z=0$, the total IR luminosity function has a significant decline in luminosity and density, but this evolution is mostly driven by the star-formation contribution rather than AGN\cite{fu10,wu11}.

Many IR galaxies at $z<1$ are in fact Seyfert galaxies, in which the total IR luminosity is mostly produced by the dust in the ISM of the host galaxies. In fact, general studies of LIRG morphologies at $z<1$ have revealed  that around half of them are spiral galaxies\cite{mel05,elb07} which contain most of the dust in their disks rather than localised in a central region.  A similar conclusion can be extracted from the IR SEDs of Seyfert galaxies: almost have of them show the signatures of cold dust and PAH emission features\cite{buc06,ram07,gal10}.

Note that mid-IR-selected AGN at $z<1$ are a distinct population from radio-loud AGN, which mainly reside in massive elliptical galaxies that are very faint (or undetected) at mid-IR wavelengths. This is consistent with the radiative-mode to radio-mode conversion scenario:  the IR-bright and radio-loud phases occur at different evolutionary stages in galaxy evolution.

\subsubsection{AGN at the Peak Activity Epoch ($1<z<3$)}

In the $\sim 4 \, \rm Gyr$ period elapsed between redshifts $z=3$ and $z=1$, the Universe had a peak in both star formation and nuclear activity\cite{hop06,ale08}. This conclusion is the result of multiple, multi-wavelength galaxy studies conducted over the last two decades. However, the exact location of this peak, or whether this maximum activity has been sustained over a significant fraction of this period, is still unknown. Only very large galaxy surveys, which allow for a detailed study of narrow redshift slices within $1<z<3$,  can clarify this issue.

From the IR perspective, much effort has been devoted to search for AGN at $1<z<3$, especially Compton-thick AGN, which are mostly undetected in large-area, X-ray surveys with moderate depths. Still today, ultra-deep X-ray data are only available for relatively small areas of the sky\cite{ale03,xue11}, so the analysis of IR galaxy surveys offers a suitable alternative to achieve a more complete census of dust-obscured AGN. 

The excellent performance of the Multiband Imaging Photometer for {\em Spitzer} (MIPS)\cite{rie04} has enabled very important progress in this subject over the last decade, as it has opened up the possibility of systematically studying mid-IR galaxies at high $z$. A notable example in the search for high-$z$ AGN candidates is the study of  bright mid-IR galaxies ($S_\nu(24 \, \rm \mu m) \gsim 300 \, \rm \mu Jy$) with high mid-IR-to-optical flux density ratios $S_\nu(24 \, \rm \mu m) / S_\nu(R) \gsim 1000$, which are commonly referred as {\em dust-obscured galaxies (DOGs)} \cite{dey08,yan04}. Different spectroscopic studies have confirmed that these galaxies mostly lie at $z\sim2$ and host AGN\cite{hou05,wee06,saj07,des08,pop08a}. This red galaxy population constitutes around a half of the most luminous ULIRGs at $z\sim2$, which indicates that nuclear activity was very common in the high $z$ ULIRG population. In turn, ULIRGs constitute $\sim 30-50\%$ of the most massive galaxies at $z\sim2-3$\cite{dad05,cap06a}. Therefore, all these results taken together indicate the importance of obscured AGN activity among massive galaxies at high $z$.

Mid-IR spectroscopy has also been very useful to unveil the presence of AGN in IR galaxies at $1<z<3$ in general, independently of their colours. The power-law shape continuum, strong silicate absorption and weak PAH emission indicate an AGN presence\cite{yan05,bra08,fad10}. For the brightest sub-millimetre-selected galaxies, {\em Spitzer} spectroscopy has shown that most of them do contain an AGN, although their total IR luminosity is dominated by star formation in most cases\cite{pop08b}. A similar conclusion was previously reached from the study of the X-ray properties of sub-millimetre sources\cite{ale05}. However, the incidence of AGN among sub-millimetre galaxies is still under debate, as other authors have found a much smaller AGN incidence among them\cite{lai10,geo11}.

In order to circumvent the limited depth of typical X-ray surveys, some authors have applied an X-ray stacking analysis of IR sources that do not show individually an AGN signature\cite{dad07,fio09}. These studies have concluded, in a statistical manner, on the presence of the elusive Compton-thick AGN population that has been predicted by models of the X-ray background\cite{gil07}.  However, some recent studies based on deep X-ray observations and/or IR spectroscopic have failed to reproduce this result through the detection of individual objects\cite{fad10,geo10}.

\subsubsection{AGN over the First Few Billion Years}

Given the sensitivity of current IR telescopes, the global population of mid-/far-IR galaxies is unknown at $z>3$ , and consequently the general incidence of AGN in IR sources at $z>3$ is a largely unexplored domain. There are some exceptions, though. In particular, sub-/millimetre-selected galaxies have a significant tail at $z>3$\cite{war11,mic12}, but the incidence of AGN among them appears to be relatively small\cite{yun12}. At these redshifts, the {\em Spitzer} IRAC mid-IR photometry appears to be essential to derive proper stellar masses, as the rest-frame near-IR light is shifted beyond $\lambda \approx 4 \, \rm \mu m$. In addition, some studies have recently identified a population of bright {\em Spitzer} sources that are very faint at near-IR wavelengths\cite{hua11,cap12,wah12}. A large fraction of these galaxies ($\gsim 50\%$) appear to be at $z>3$, and the brightest ones show signs of AGN activity\cite{cap14}. 

Most  AGN IR studies at $z>3$ correspond to the follow up of AGN selected at other wavelengths. This is particularly the case of high-$z$, luminous radiogalaxies. {\em Spitzer} observations of these sources clearly show the power-law signature of AGN in their mid-IR SEDs\cite{fra04,sey07,deb10,roc13}.   On the other hand, IR observations of bright QSOs at $z\gsim 5$ have revealed the presence of hot and cold dust in these objects, and also CO emission, implying the existence of large molecular gas reservoirs\cite{car07,wan08,wan10, wan11,gil14}. I have discussed recent results on this subject in section \S\ref{sec_agnsfr}. However, not all known quasars at $z\sim6$ appear to be bright at sub-/millimetre wavelengths\cite{omo13}. This result is in line with the latest studies of optically-selected, high-$z$ QSOs at mid-IR wavelengths. Blain et al. (Ref.\refcite{bla13}) presented a {\em WISE} mid-IR data analysis for 31 QSOs at $z>6$. They found that half of them were individually detected by {\em WISE}, while the other half was not. A stacking analysis of this second half reveals that they are considerably fainter at mid-IR wavelengths than the brighter half. The cause of this possible dichotomy remains unclear, but it is possible that these QSOs are at different stages of their accretion histories, and in some of them powerful outflows have swept the dust away from their host galaxies.

\subsubsection{AGN in Infrared Galaxy Models}

Recent IR galaxy models have incorporated AGN, mostly in an empirical way\cite{fra08,val09,gru11}, but a few of them have done it using theoretical arguments\cite{cai13}. Although considering AGN does not appear to be mandatory to reproduce the galaxy number counts at IR wavelengths, it is  necessary to reproduce other observables, such as the bright-end of the IR galaxy luminosity function. Moreover,  reproducing the IR galaxy redshift distributions may require some evolution in the AGN contribution to IR sources with redshift\cite{val09}.

Other works have been able to predict different IR galaxy observables after incorporating the effects of AGN feedback within the framework of cold dark matter galaxy models\cite{lac08,hop10,som12,hay13}. However, some of them cannot reproduce the bright end of the IR galaxy luminosity function at $z\sim2$, presumably for not including sufficient numbers of obscured AGN\cite{som12}. Some models have managed to reproduce this bright end, but they need to invoke a top-heavy IMF\cite{bau05,lac08}.

Recently, Hopkins et al. (Ref.~\refcite{hop10}) have predicted the contribution of AGN to the total IR galaxy luminosity function through cosmic time, based on an empirical halo occupation model coupled with hydrodynamical simulations. AGN, along with galaxy mergers, dominate the bright end of this luminosity function, in agreement with observational results\cite{bab06,gru10}, while galaxies with more regular star formation activity prevail at lower luminosities, up to $z\sim3$. At higher redshifts, the different components may contribute comparably to the bright end, but uncertainties are large. Overall, the AGN contribution to the IR luminosity function is of a few percent at all redshifts, and follows the same trend as all other galaxy types:  the associated IR luminosity density remains almost constant between $z=4$ and $z=1$, and then has a quick decline to $z=0$.

\subsubsection{Constraints from/to the Cosmic Infrared Background}

The total Cosmic Infrared Background (CIB) and its evolution with redshift have been quantified by different studies using data from the latest IR telescopes\cite{dol06,mar09,ber10,jau11}. Understanding the contribution of AGN to the overall IR background at different redshifts is, however,  a difficult task, because AGN identification remains quite incomplete in large areas of the sky, for which only datasets of moderate depths are available. 

In spite of this difficulty, some authors have attempted to obtain estimates of the AGN contribution to the CIB.  Serjeant et al. (Ref.\refcite{ser10}) carried out a joint analysis of AKARI and SCUBA2 data, and derived that $15 \, \rm \mu m$ sources with $S_\nu(15 \, \rm \mu m)> 200 \, \rm \mu Jy$ make more than 10\% of the $450 \, \rm \mu m$ background, but less than 4\% of the 1.1~mm background light. Cappelluti et al. (Ref.\refcite{cpp13}) performed a cross-correlation analysis of the CIB fluctuations at 3.6 and 4.5~$\rm \mu m$ and X-ray fluctuations at 0.5-7~keV, and concluded that 15-25\% of the unresolved CIB at these wavelengths is produced by an unidentified population of X-ray emitters at 0.5-2.0~keV, probably obscured AGN at $z>2$. 

On the other hand, Shi et al. (Ref.\refcite{shi13a}) estimated the abundance of Compton-thick AGN based on a joint model of the X-ray and IR backgrounds\cite{shi13b}. They found that the fraction of Compton-thick AGN to all AGN is of only a few percent at  $f(2-10 \,\rm keV)>10^{-15} \, \rm erg \, s^{-1} cm^{-2}$, but increases  at fainter fluxes, and contributes as much as 25\% to the total X-ray background at 20~keV. These authors argued that around 90\% of the Compton-thick AGN should be detectable at $5-10 \, \rm \mu m$ due to the hot dust emission of their dusty tori.

Consistently with other works based on the study of resolved IR galaxy populations,  these results indicate that AGN make a relatively small fraction of the total CIB energy budget. This is probably the case even considering the fraction of the CIB which is still not accounted for in current IR galaxy surveys. Taking into account that host galaxy stellar masses and black-hole masses differ in a factor of $\sim100$, it is perhaps not surprising that most of the energy budget is in star formation rather than black-hole activity.

\section{Future Prospects}
\label{sec_fut}

The wealth of results discussed here show the significant progress made over the last 10-15 years towards the understanding of AGN properties, and the overall role of AGN in galaxy evolution, thanks to the analysis and modelling of IR astronomical data. Many questions remain unanswered, though. I identify here some of these main open issues, and  briefly discuss the prospects for tackling them, making use of the main IR astronomical facilities that are forthcoming within the next decade or so.

1.{\em The general existence and geometry of the dusty torus}.  Systematic, high-spatial resolution studies able to resolve the inner pc scale of AGN in the local Universe are necessary to test the fundamental prediction of the unification models, i.e. the general existence of the dusty torus. They should also reveal whether the size and geometry of this torus vary with the black-hole mass and/or accretion power. Moreover, it would be relevant to investigate  the relation between the properties of the dusty torus and the presence of gas outflows originated at the central engine. Could these outflows modify, or even destroy, the dusty torus? Sub-arcsec resolution images  taken with the full ALMA array at sub-/millimetre wavelengths,  in the mid-IR regime with the Mid Infrared Instrument (MIRI)\cite{wri08} on the {\em James Webb Space Telescope (JWST)}, and later the METIS\cite{bra12} instrument on the Extremely Large Telescope (ELT), should enable a systematic investigation of these problems.

2.{\em The possible mechanisms for black-hole formation, and the onset of a black-hole mass/galaxy bulge mass correlation in cosmic time}. The universality of a black-hole/host galaxy relation is currently under scrutiny, as recent observational results indicate that black holes may reside in galaxies of different morphological types. Even if such a correlation holds for galaxy bulges,  this correlation may not have been in place until relatively late in cosmic time ($z\sim1$).  This would mean that the prevalent mechanisms for black-hole fuelling  could have changed through cosmic history. However, an important caveat of current observations are morphological k-corrections: to properly compare the morphologies of AGN hosts at different redshifts, high resolution images at similar {\em rest-frame}  near-IR wavelengths (relatively unaffected by dust obscuration) are needed. This is another area in which forthcoming {\em JWST}  images will play a fundamental role to clarify the picture.

3.{\em A (complete) census of the AGN incidence among galaxies of different stellar masses through cosmic time}. For an ultimate understanding of the role of AGN in galaxy evolution, it is necessary to conduct a large and unbiased AGN census, and determine where these AGN live at different redshifts. The IR regime is ideal to provide an efficient diagnostic of the AGN presence in large galaxy samples. High-sensitivity, wide-field telescopes providing spectroscopy to near-IR wavelengths, such as the {\em Euclid} mission, should allow for a simultaneous selection of stellar-mass-selected galaxy samples, and AGN identification, over large areas of the sky. Given the covered wavelength range, these galaxy samples in stellar mass should be quite complete up to $z\sim2$. Although stellar mass completeness will progressively be lost at higher redshifts, {\em Euclid} will still allow for a large census of the less obscured AGN up to $z\sim6-7$.

4.{\em The radiative-mode / radio-mode evolutionary sequence}. Understanding whether this path is universal for all AGN requires a detailed determination of the number density evolution of X-ray-selected and radio-selected AGN, through cosmic time. This is a major topic that will benefit from the synergies between wide-field near-IR telescopes, like {\em Euclid}, and major radio-telescopes (e.g., LOFAR and later the SKA).

5.{\em  The appearance of the first AGN in the early Universe.} Until now, only very luminous quasars are known at $z>4$, with only a handful of cases at redshifts $z>6$.  Investigating the evolution of the bulk of the AGN population and their hosts at these high redshifts is fundamental to constrain galaxy formation models. In addition, the study of the first appearance of AGN at the epoch of reionisation should clarify whether these objects had any significant role in this process (which so far is suspected to be minor).  The deepest {\em JWST} images will constitute a unique treasure trove for this purpose.

\section*{Acknowledgments}

The author is grateful to Almudena Alonso-Herrero, Peter Barthel and Roberto Maiolino for useful discussions and comments on this manuscript; and to Dale Kocevski and Roberto Maiolino for providing figures.

\bibliographystyle{ws-ijmpd_caputi_1authetal}
\bibliography{biblio_agnirrev}

\end{document}